\documentclass[11pt]{article}
\usepackage{graphicx}
\usepackage{epsfig}

\newlength{\bredde}
\def\slash#1{\settowidth{\bredde}{$#1$}\ifmmode\,\raisebox{.15ex}{/}
\hspace*{-\bredde} #1\else$\,\raisebox{.15ex}{/}\hspace*{-\bredde} #1$\fi}
\newcommand{\sect}[1]{\setcounter{equation}{0}\section{#1}}

\newcommand{\beq}{\begin{equation}}
\newcommand{\eeq}{\end{equation}}
\newcommand{\ba}{\begin{array}{ccc}}
\newcommand{\ea}{\end{array}}
\newcommand{\nn}{\nonumber}

\def\beqn{\begin{eqnarray}}
\def\eeqn{\end{eqnarray}}

\def\hx{\hat{x}}
\def\hy{\hat{y}}
\def\hm{\hat{m}}
\def\hmu{\hat{\mu}}
\def\hd{\hat{\delta}}
\def\hchi{\hat{\chi}}
\def\hH{\hat{H}}

\def\Sl#1{\rlap{\raisebox{.15ex}{$\mskip 4 mu /$}}#1}  
\begin{document}
\topmargin -1.4cm
\oddsidemargin -0.8cm
\evensidemargin -0.8cm
\title{\Large{\bf 
A new Chiral Two-Matrix Theory for 
Dirac Spectra with Imaginary Chemical Potential
}}

\vspace{2cm}

\author{~\\{\sc G. Akemann}$^1$, ~~{\sc P.H. Damgaard}$^2$,  
~~{\sc   J.C. Osborn}$^3$ ~~and~~ {\sc K. Splittorff}$^2$
\\~\\~\\
$^1$Department of Mathematical Sciences\\
Brunel University West London\\ 
Uxbridge UB8 3PH\\ United Kingdom
\\~\\~\\
$^2$The Niels Bohr Institute\\ Blegdamsvej 17\\ DK-2100 Copenhagen {\O}\\
Denmark
\\~\\~\\
$^3$Physics Department \& Center for Computational Science\\
Boston University\\
Boston, MA 02215\\
USA
}
\date{}
\maketitle
\vfill
\begin{abstract}
We solve a new chiral Random Two-Matrix Theory 
by means of biorthogonal polynomials for
any matrix size $N$. By deriving the relevant kernels we
find explicit formulas for all $(n,k)$-point spectral (mixed or unmixed)
correlation functions. In the microscopic limit we find the corresponding scaling
functions, and thus derive all spectral correlators in this limit as well.
We extend these results to the ordinary (non-chiral) ensembles, and also there provide
explicit solutions for any finite size $N$, and in the microscopic scaling
limit. Our results give the general analytical expressions for the
microscopic correlation functions of the Dirac operator eigenvalues in theories with
imaginary baryon and isospin chemical potential, and can be used to extract
the tree-level pion decay constant from lattice gauge theory configurations.
We find exact agreement with previous computations based on the low-energy
effective field theory in the two
special cases where comparisons are possible.
\end{abstract}
\vfill

\newpage


\sect{Introduction}

In this paper we propose a chiral Random Matrix Theory of two coupled hermitian
matrices, and solve the theory exactly at finite matrix size $N$. In the $N
\to \infty$ limit we find the associated microscopic scaling regime, and
derive closed expressions for all spectral correlation functions. It has been
shown recently \cite{DHSS,DHSST} that these spectral correlation functions
describe correlations between the lowest-lying Dirac operator eigenvalues in
the $\beta=2$ Dyson universality class of spontaneous chiral symmetry
breaking \cite{Jac0}, once subjected to an external vector field. The effect 
of the external vector source on the chiral condensate was considered in ref. 
\cite{MT} and very recently spectral sum rules have been derived
directly from the effective partition functions \cite{L}. The chiral
Random Two-Matrix Theory generalizes the original one-matrix problem of chiral
Random Matrix Theory \cite{SV,VZ,ADMN,DN} (the microscopic hard-edge
problem with a certain determinant measure)
and is also related to the chiral non-Hermitian Random Two-Matrix Theory
\cite{O} for QCD with a chemical potential by setting the chemical potential
to be purely imaginary.

The solution of the chiral Random Two-Matrix Theory is of interest in its
own right, but it is also of importance in the context of lattice gauge 
theory. This is due to the fact that the spectral correlation functions of
the chiral two-matrix problem can be used to measure the pion decay constant
$F_{\pi}$ in a manner analogous to how the chiral condensate $\Sigma$ can be
extracted by means of the original microscopic solution of the chiral
Random Matrix Theory of just one chiral matrix \cite{DHK}. The idea is to 
compare the analytic finite-volume scaling functions of Dirac operator
eigenvalues to lattice simulation data. Fits to these data yield
accurate determinations of the two low-energy constants of QCD, $\Sigma$ and
$F_{\pi}$.

Before we describe the relevant chiral Random Two-Matrix Theory, we will
briefly outline how this two-matrix problem arises in the context of particle
physics and how $F_\pi$ enters the game. The coupled chiral two-matrix
problem turns out to describe the microscopic limit of QCD Dirac operator 
spectra close to the origin. 
The inclusion of one additional matrix in the way we shall describe below
allows us to determine the correlations between Dirac operators with  
different imaginary chemical potentials. In detail, we shall consider
two different Dirac operators, 
\beqn
D_1\psi_1^{(n)} &\equiv & [\Sl{D}(A)+i\mu_1\gamma_0]\psi_1^{(n)} ~=~
 i\lambda_1^{(n)}\psi_1^{(n)}  \cr
D_2\psi_2^{(n)} &\equiv & [\Sl{D}(A)+i\mu_2\gamma_0]\psi_2^{(n)} ~=~
 i\lambda_2^{(n)}\psi_2^{(n)} \ .
\label{Ddef}
\eeqn
The interest in studying these Dirac operators has already been
hinted at above: The correlation functions between eigenvalues of Dirac
operators at different imaginary chemical potential is extremely sensitive to
the value of $F_\pi$. In the so-called $\epsilon$-regime of QCD this
dependence on $F_\pi$ is uniquely determined by the pattern of chiral
symmetry breaking.  
As we will show below the microscopic correlations between the two different
sets of eigenvalues $\lambda_1$ and $\lambda_2$ only depends on the 
difference $\mu_2-\mu_1$.
One can then for convenience set $\mu_2 = -\mu_1 = \mu$
and study the theory with an imaginary isospin chemical potential $\mu$.
Alternatively we will also consider the case where $\mu_1=0$ and
$\mu_2 \ne 0$.
Then $D_1$ is the usual Dirac operator of a given gauge
configuration $A_{\mu}(x)$ and is what is typically used in lattice
QCD simulations.
This allows one to extract the low energy constants from existing
simulations.

A powerful way to extract information about the lowest-lying Dirac
operator eigenvalues is through the effective low-energy theory of
Goldstone bosons, the associated chiral Lagrangian \cite{LS}. In the
$\epsilon$-regime this effective low-energy field theory receives, to
leading order, only contributions from momentum zero modes of the
Goldstone bosons. In addition, the topological index $\nu$ of the
given gauge field configuration now plays a crucial role, and it is
advantageous to consider the partition function in a sector of fixed
topological charge $\nu$. For a theory of $N_f$ quark flavors, we let
$B$ denote the $N_f\times N_f$ diagonal matrix of imaginary chemical
potentials,
i.e. for the two Dirac operators of eq. (\ref{Ddef})
a diagonal matrix of $\mu_1$'s and/or $\mu_2$'s. In the effective
SU($N_f$) low-energy Lagrangian the fundamental field is taken as
$U(x) \in$ SU($N_f$).  The translation table from quark couplings to
couplings in the effective theory is well known. The quark chemical
potentials is of vector-kind, and only in the time direction.  The
coupling to external imaginary chemical potentials described by the
matrix $B$ is then achieved by replacing the time-derivative,
\beq
\partial_0U(x) ~\to~ \nabla_0U(x) ~=~ \partial_0U(x) -i [B,U(x)] .
\label{Ucov}
\eeq    
In both the more conventional ``$p$-regime'' and the $\epsilon$-regime
the leading-order effective Lagrangian is
\beq
{\cal L} = \frac{F_{\pi}}{4}{\rm Tr}\left[\nabla_0U(x)\nabla_0U^\dagger(x) +
\partial_iU(x)\partial_iU^\dagger(x)\right] -
\frac{\Sigma}{2}{\rm Tr}\left[{\cal M}U^{\dagger}(x)
+ {\cal M}^{\dagger}U(x)\right], \label{L}
\eeq
where ${\cal M}$ is the (diagonal) quark mass matrix of size
$N_f\times N_f$.  This effective theory simplifies even further in the
$\epsilon$-regime, where to leading order only momentum zero modes
play a role. At fixed gauge field topology $\nu$ the partition
function then reads
\beq
{\cal Z}_{\nu} =
\int_{U(N_f)} dU \,(\det U)^{\nu}e^{\frac{1}{4}VF_\pi^2
{\rm Tr} [U,B][U^\dagger,B] + \frac{1}{2} \Sigma V
{\rm Tr}({\cal M}^\dagger U + {\cal M}U^\dagger)} ~,
\label{Zeff}
\eeq
where the integration is performed over the full unitary group. Note that
$F_\pi$ enters in the dimensionless combination $VF_\pi^2\mu_f^2$.
In the case where all chemical potentials are equal (imaginary baryon
chemical potential), $B$ is proportional to the identity and all
$\mu$ and $F_\pi$-dependence trivially drops
out of the effective low-energy Lagrangian as expected since the Goldstone
bosons do not carry baryon charge. However, when different chemical
potentials enters there is a unique dependence on $F_\pi$. The
correlation function between the spectra of $D_1$ and $D_2$ have a
dramatic dependence on $F_\pi$. In \cite{DHSS,DHSST} the two point
function was obtained for zero and 
for two flavors directly from the low-energy effective chiral
Lagrangian. Here we generalize these results using a chiral Random Matrix
Theory. The chiral Random Matrix Theory approach to the $\epsilon$-regime
in this case is a two-matrix problem. We need two matrices because we wish to
describe two different kinds of operators, $D_1$ and $D_2$. With this
extension of the original chiral Random Matrix Theory we will derive all
$(n,k)$-point correlation functions of the operators $D_1$ and $D_2$.
The advantage of the chiral Random Two-Matrix Theory approach over the direct
calculation using the low-energy effective chiral Lagrangian lies in the ease
with which the calculations generalize. We will verify explicitly that the
chiral Random Two-Matrix Theory reproduces, in two special cases, the 
results obtained previously through the low-energy effective chiral
Lagrangian \cite{DHSS,DHSST}. 

\vspace{3mm}

In this paper we will also find the corresponding solution of the non-chiral
Random Matrix Theory with two coupled matrices generalizing the results of
Eynard and Mehta \cite{EM} to include determinants (matrix analogues of Dirac
determinants). This is of relevance for
Dirac operator spectra of the analogous 3-dimensional gauge theory
\cite{VZ3,Sz}. It may have applications in condensed matter physics as well
\cite{HN}.   

\vspace{3mm}

We round off this introduction with a few remarks on the case of real
chemical potential. 
The spectral properties of the Dirac operator change drastically if we were
to consider instead ordinary real baryon chemical potential. While both
Dirac operators (\ref{Ddef}) are anti-hermitian (and all
$\lambda_i^{(n)}$ thus real), this is not the case for real chemical
potential. The eigenvalues in that case lie in the 
complex plane. The precise manner in which the eigenvalues spread into the
complex plane when a real baryon chemical potential is turned on has been
computed in the $\epsilon$-regime \cite{O} by a random matrix approach and
subsequently verified by a low-energy effective calculation \cite{AOSV}.

A Random One-Matrix Theory describing real chemical potential in the low-energy
limit was first suggested by Stephanov \cite{Steph}.
As shown in \cite{O}, the way to find an appropriate eigenvalue representation
is to start out with a {\em two-matrix} extension, of which one matrix is
anti-hermitian, and the other is hermitian. The eigenvalues of the combination
are complex, and their microscopic distributions coincide with the low-energy
Dirac operator
eigenvalues at real baryon chemical potential. Much work has recently
gone into the understanding of this low-lying Dirac spectrum 
\cite{G1,AV,SplitV,O,AFV,AOSV,G2} (for numerical evidence from lattice gauge
theory simulations see, $e.g.$, ref. \cite{TO}).
As stated earlier the Random Matrix Theory at real baryon chemical potential
introduced in \cite{O} is related to the chiral Random Two-Matrix Theory
for imaginary chemical potential we propose in this paper.
We will point out the similarities and differences below.

Our paper is organized as follows. In the next section we introduce
the chiral Random Two-Matrix Theory, and proceed to derive an
eigenvalue representation.  This leads us to the use of biorthogonal
polynomials, which we identify explicitly for finite size of the
matrices in section \ref{sec-expcomp}.
There we present our results for all spectral
correlation functions in both the quenched case, the partially
quenched case, and the full case (appropriate definitions of these
terms will be given below).  We also derive the microscopic scaling
limit in all three cases.  In section \ref{sec-nonchiral}
we turn to the non-chiral
variant of this theory, and find also there the explicit solution both
at finite $N$ and in the analogous scaling limit.  Section \ref{sec-summary}
contains our conclusions. Several technical details are collected in 3
Appendices. 


\sect{The Chiral Two-Matrix Problem}\label{ch2MM}

The chiral Two-Matrix Theory with imaginary chemical potential is described
by the partition function 
\beqn
{\cal Z}_{\nu}^{(N_f)}(\{m\}) &\sim&
 \int d\Phi  d\Psi~ \exp\left[-{N}{\rm Tr}\left(\Phi^{\dagger}
\Phi + \Psi^{\dagger}\Psi\right)\right]  
\prod_{f=1}^{N_f} \det[{\cal D}(\mu_f) + m_{f}] 
\label{ZNf}
\eeqn
where the anti-Hermitian operator ${\cal D}$ in the determinant is given by
\beqn
{\mathcal D}(\mu_f) = \left( \begin{array}{cc}
0 & i \Phi + i \mu_f \Psi \\
i \Phi^{\dagger} + i \mu_f \Psi^{\dagger} & 0
\end{array} \right) ~.
\eeqn
Both $\Phi$ and $\Psi$ are complex rectangular matrices of size
$N\times (N+\nu)$, where $N$ and $\nu$ are integers.
As shown in (\ref{ZNf}) we restrict ourselves to Gaussian weight functions
for both $\Phi$ and $\Psi$ in this paper. 
The partition function \cite{O} at real chemical potential is obtained by
taking all $\mu_f$ purely imaginary. Because this changes the Hermiticity 
properties of the Dirac operator the eigenvalue correlation functions at real
and imaginary chemical potential are {\sl not} related by a simple analytic
continuation $\mu_f\to i\mu_f$.

We stress already here that $\mu_f$ and $m_f$ are (dimensionless)
quantities that a priori have no relation to either chemical potential
or quark masses in QCD. However, in the large-$N$ scaling limit that
we shall discuss below they do translate into the familiar low-energy
parameters of QCD by means of $\hat{\mu}_f^2 = 2\mu_f^2 N \to
\mu_f^2F_{\pi}^2V$ and $\hat{m} = 2mN \to m\Sigma V$. With this in mind
we refer to $N_f$ as the number of flavors (here we consider only
$N_f\geq 0$).

We will consider the eigenvalues of the two matrices
$\mathcal{D}_1 \equiv \mathcal{D}(\mu_1)$ and
$\mathcal{D}_2 \equiv \mathcal{D}(\mu_2)$ for
arbitrary (real) values of $\mu_1$ and $\mu_2$.
The case of imaginary isospin chemical potential can be obtained from this
by setting $\mu_2 = -\mu_1 = \mu$.
To obtain the eigenvalues we first change integration variables to
\beqn
\Phi_1 &\equiv& \Phi + \mu_1 \Psi \cr
\Phi_2 &\equiv& \Phi + \mu_2 \Psi ~. \label{Phipm}
\eeqn
We will restrict ourselves to the case where all determinants in (\ref{ZNf})
contain $\mathcal{D}_1$ or $\mathcal{D}_2$.
Their numbers are denoted by $N_1$ and $N_2$,
 with masses $m_{f1}$ and $m_{f2}$, respectively:
\beqn
{\cal Z}_{\nu}^{(N_f)}(\{m_1\},\{m_2\}) \sim
 \int\! d\Phi  d\Psi 
\prod_{f1=1}^{N_1}\! \det[{\cal D}_1 + m_{f1}]\!
\prod_{f2=1}^{N_2} \!
\det[{\cal D}_2 + m_{f2}]\ 
e^{-N{\rm Tr}\left(\Phi^{\dagger}\Phi+\Psi^{\dagger}\Psi\right)}
\label{ZNfiso}
\eeqn
with $N_f=N_1 +N_2$.
After the change of variables (\ref{Phipm}) the partition function becomes
\beqn
{\cal Z}_{\nu}^{(N_f)}(\{m_1\},\{m_2\}) \sim
 \int \!\! d\Phi_1  d\Phi_2 \! \prod_{f1=1}^{N_1}\! \det[{\cal D}_1 + m_{f1}]
\prod_{f2=1}^{N_2}\! 
\det[{\cal D}_2 + m_{f2}]\ 
e^{-N{\rm Tr}V(\Phi_1,\Phi_2)}
\label{ZNfpm}
\eeqn
where each of the two operators are of the standard chiral form,
\beqn
\mathcal{D}_1 = \left( \begin{array}{cc}
0 & i \Phi_1 \\
i \Phi_{1}^{\dagger} & 0
\end{array} \right) 
\eeqn
and 
\beqn
\mathcal{D}_2 = \left( \begin{array}{cc}
0 & i \Phi_2 \\
i \Phi_{2}^{\dagger} & 0
\end{array} \right) ~.
\eeqn
However the two matrices $\Phi_1$ and $\Phi_2$ are now coupled:
\beq
V(\Phi_1,\Phi_2) =
  c_1 \Phi_{1}^{\dagger}\Phi_1 + c_2 \Phi_{2}^{\dagger}\Phi_2
- d \left(\Phi_{1}^{\dagger}\Phi_2
+\Phi_{2}^{\dagger}\Phi_1\right)
\label{V}
\eeq
with the abbreviations 
\beqn
c_1 &=& (1+\mu_2^2)/\delta^2  \cr
c_2 &=& (1+\mu_1^2)/\delta^2 \cr
d &=& (1+\mu_1\mu_2)/\delta^2 \cr 
\delta &=& \mu_2 - \mu_1 \ ,
\eeqn
used in the following.
This is the chiral analogue of the Two-Matrix problem introduced in
the classic papers of refs. \cite{IZ,M}. We solve this chiral theory
by the biorthogonal polynomial method, analogous to what is described
in ref. \cite{M}.  At this stage, {\it after} changing to variables
$\Phi_{1,2}$ we could in principle allow for the following more
general class of weight functions:\footnote{Switching to higher order
polynomials before the change of variables leads to higher order terms
in $\Phi_{1}^{\dagger}\Phi_2$ and its conjugate.}
\beq
V(\Phi_1,\Phi_2) \to v_1\left(\Phi_{1}^{\dagger}\Phi_1\right)
+v_2\left(\Phi_{2}^{\dagger}\Phi_2\right) - d \left(\Phi_{1}^{\dagger}\Phi_2
+\Phi_{2}^{\dagger}\Phi_1\right) ~,
\eeq
where $v_{1,2}$ are polynomials.
It remains an interesting open challenge to establish universality of
the present results in the large-$N$ limit. A generalization of the
universality proof in \cite{ADMN,DN} is non-trivial due to the
different structure of recursion relations for biorthogonal
polynomials in the two-matrix theory.

The first step now consists in finding an eigenvalue representation for the
partition function (\ref{ZNfpm}). To this end we follow the standard procedure \cite{SV}
and write
\beqn
\Phi_1 &=& U_1XV_{1}^{\dagger} \cr
\Phi_2 &=& U_2YV_{2}^{\dagger} ~,
\eeqn
where $U_{1,2}$ and $V_{1,2}$ are unitary matrices (with $V_{1,2}$ restricted
to $U(N+\nu)/[U(1)]^{N+\nu}$ \cite{VZ}), and $X$ and $Y$ are diagonal matrices
with real and non-negative entries $x_i$ respectively $y_i$. Under this
change of variables the measure (including the determinants) becomes
\beqn
d\Phi_1 \prod_{f1=1}^{N_1}\! \det[{\cal D}_1 + m_{f1}] &=& dU_1 dV_1 \prod_{i=1}^N \left(dx_i^2 x_i^{2\nu}
\prod_{f1=1}^{N_1} (x_i^{2} + m_{f1}^2)\right) \Delta_N(\{x^2\})^2 \cr
d\Phi_2 \prod_{f2=1}^{N_2}\! 
\det[{\cal D}_2 + m_{f2}]&=& dU_2 dV_2 \prod_{i=1}^N \left(dy_i^2 y_i^{2\nu}
\prod_{f2=1}^{N_2} (y_i^{2} + m_{f2}^2)\right) \Delta_N(\{y^2\})^2
\eeqn
where we have left out the factors of $m_{f}^\nu$ coming from the zero
modes of the determinants.
Here $U_{1,2}$ and $V_{1,2}$ are integrated over the Haar measure, 
and $\Delta_N$ is the Vandermonde determinant. Because of the coupling between
$\Phi_1$ and $\Phi_2$ the integration over $U_{1,2}$ and $V_{1,2}$ is not
trivial\footnote{Of course this is only true for $\mu_{1,2}\neq0$.}. Using
left and right invariance of the Haar measure the required integral reduces to a form
that has been evaluated in ref. \cite{JV}:
\beq
\int dUdV \exp\left[d N {\rm Tr}(VXUY) ~~{\rm + ~h.c.}~ \right] 
~=~ \frac{\det[I_{\nu}(2 d N x_iy_j)]}
{\prod(x_iy_i)^{\nu}\Delta_N(\{x^2\})\Delta_N(\{y^2\})} ~,
\eeq
up to an irrelevant normalization factor. Here $I_\nu(x)$ is a modified Bessel
function. 

Putting everything together we thus finally arrive at the eigenvalue representation
\beqn
{\cal Z}_{\nu}^{(N_f)}(\{m_1\},\{m_2\}) &=& {\cal N}(\{m_{1,2}\}) \int_0^{\infty} \prod_{i=1}^N\left(dx_idy_i (x_iy_i)^{\nu+1}
\prod_{f1=1}^{N_1} (x_i^2+m_{f1}^2)
\prod_{f2=1}^{N_2} (y_i^2+m_{f2}^2) \right) \cr
&\times&\Delta_N(\{x^2\})\Delta_N(\{y^2\})\det\left[I_{\nu}(2 d N x_i y_j)\right] 
e^{-N \sum_i^N c_1 x_i^2 + c_2 y_i^2 } \label{evrep}
\eeqn
where the factors of $m_{f}^\nu$ from the zero modes have now been absorbed
in the normalization constant ${\cal N}$.
In contrast to the one-matrix theory we cannot extend the integration
range to $-\infty$, as the integrand is no longer even in $x_i$ and $y_i$.
The resulting expression (\ref{evrep}) has a quite striking resemblance to the non-hermitian
theory describing the Dirac operator spectrum in the presence of
real baryon chemical potential \cite{O}. In that case the explicit
Bessel function in the measure is of the kind $K_{\nu}$ rather than the $I_{\nu}$
of eq. (\ref{evrep}).
In the eigenvalue representation (\ref{evrep}) 
we can apply the technique of biorthogonal polynomials. 
For that purpose we seek
two sets of monic polynomials $P_n^{(N_f)}(x^2)$ and $Q_n^{(N_f)}(y^2)$ 
that are biorthogonal,
\beq
\int_0^{\infty} dx dy \ w^{(N_f)}(x,y)\ 
P_n^{(N_f)}(x^2)\ 
Q_k^{(N_f)}(y^2)
= h_n^{(N_f)} \delta_{nk} ~, \label{bio}
\eeq
with respect to the weight function
\beq
w^{(N_f)}(x,y)~\equiv~ 
(xy)^{\nu+1}\prod_{f1=1}^{N_1}(x^2+m_{f1}^2)
\prod_{f2=1}^{N_2}(y^2+{m}_{f2}^2)\ 
I_{\nu}\left(2dNxy\right) e^{-N(c_1 x^2+ c_2 y^2)}\ .
\label{weight}
\eeq
Note that the order of arguments in $w^{(N_f)}(x,y)$ is important, as
$x$ only appears in the $N_1$ mass terms, while $y$ only appears in the
$N_2$ mass terms.
Only in the special case of $\mu_2 = -\mu_1$, $N_1=N_2$ and masses  pairwise
matched, $m_{f1} = m_{f2}$, is the measure symmetric in $x$ and $y$.
This is in particular true for the quenched case with an imaginary isospin
chemical potential.

After the usual rewriting of the Vandermonde determinants,
\beqn
\Delta_N(\{x^2\}) &=& \det\left[x_i^{2(j-1)}\right] ~=~ \det\left[P_{j-1}^{(N_f)}(x_i^2)\right] \cr
\Delta_N(\{y^2\}) &=& \det\left[y_i^{2(j-1)}\right] ~=~ \det\left[Q_{j-1}^{(N_f)}(y_i^2)\right] 
\eeqn
in terms of polynomials in squared variables 
one can use the biorthogonality condition (\ref{bio}) to perform the integral
(\ref{evrep}) in terms of the norms  $h_n^{(N_f)}$ giving
${\cal  Z}_{\nu}^{(N_f)}=N!\prod_{n=0}^{N-1}h_n^{(N_f)}$.  
The result can be expressed in terms of the kernel $K_N(x,y)$ of the quenched polynomials
$P_k^{(0)}(x^2)$ and $Q_k^{(0)}(y^2)$, defined in eq. (\ref{K}) below. 
Details of the derivation following \cite{AV} are given in
appendix \ref{allZ} leading to 
\beqn
\label{Zgen}
{\cal Z}_{\nu}^{(N_f)}(\{m_1\},\{m_2\}) &=& N!\prod_{f1=1}^{N_1}m_{f1}^\nu\prod_{f2=1}^{N_2}m_{f2}^\nu
\frac{\prod_{j=0}^{N+N_2-1}h_j^{(0)}}{\Delta_{N_1}(\{m_1^2\})\Delta_{N_2}(\{m_2^2\})}\\
&\times&
\det\left[ K_{N+N_2}^{(0)}(im_{k1},im_{l2}) \ P_{N+N_2}^{(0)}((im_{k1})^2)\ldots  
P_{N+N_1-1}^{(0)}((im_{k1})^2)\right],
\nn
\eeqn
for $N_1\geq N_2$. Here we have specified the normalization leading to 
${\cal N}(\{m_{1,2}\})=\prod_{f1=1}^{N_1}m_{f1}^\nu\prod_{f2=1}^{N_2}m_{f2}^\nu$.

We are not only interested in the partition function itself, but mostly in the spectral correlation
functions. The biorthogonal polynomial technique has been adopted to this purpose in
refs. \cite{MS,EM,E}. We review it briefly here, and in particular we make use
of the very compact 
and general formulation of \cite{EM}. The measure considered in that
paper is of the ordinary 
unitary theory which through the Itzykson-Zuber integral \cite{IZ} gives rise to an
exponential coupling between the two sets of eigenvalues. In our case this is replaced
by the coupling through the Bessel function $I_{\nu}$ as in eq. (\ref{bio}). However one
can straightforwardly follow through the same steps as in \cite{EM} with this
replacement. In general four different kernels are required.
Three of the kernels also depend on the generalized Bessel transforms
\beqn
\chi_k^{(N_f)}(y) &\equiv& \int_0^{\infty}dx\ w^{(N_f)}(x,y) P_k^{(N_f)}(x^2) \cr
\hchi_k^{(N_f)}(x) &\equiv& \int_0^{\infty}dy\ w^{(N_f)}(x,y) Q_k^{(N_f)}(y^2) 
~.\label{chidef}
\eeqn
Note that in general the two functions 
are not polynomials. In
particular, they are proportional to the part of the weight which is not integrated.

The four required kernels in terms of the biorthogonal polynomials and their
transforms are:
\beqn
K_N^{(N_f)}(y,x) 
&=& 
\sum_{k=0}^{N-1}\frac{Q_k^{(N_f)}(y^2)P_k^{(N_f)}(x^2)}{h_k}\ ,
\label{K}\\
H_N^{(N_f)}(x_1,x_2) &=&
\sum_{k=0}^{N-1}\frac{\hchi_k^{(N_f)}(x_1)P_k^{(N_f)}(x_2^2)}{h_k} 
~=~ \int_0^\infty dy\  w^{(N_f)}(x_1,y)K_N^{(N_f)}(y,x_2)
\label{H}
\\
\hH_N^{(N_f)}(y_1,y_2) &=&
\sum_{k=0}^{N-1}\frac{Q_k^{(N_f)}(y_1^2)\chi_k^{(N_f)}(y_2)}{h_k} 
~=~ \int_0^\infty dx\ w^{(N_f)}(x,y_2)K_N^{(N_f)}(y_1,x)
\label{Hh}
\\
M_N^{(N_f)}(x,y) &=& 
\sum_{k=0}^{N-1}\frac{\hchi_k^{(N_f)}(x)\chi_k^{(N_f)}(y)}{h_k}
~=~ \int_0^\infty dudv\ w^{(N_f)}(x,u) w^{(N_f)}(v,y)  K_N^{(N_f)}(u,v).
\label{M}
\eeqn
Note that also here the order of arguments is important. We have chosen the notation 
$x_i$ and $y_i$ to indicate the nature of the arguments so that $x_i$ is an eigenvalue
of associated with ${\cal D}_1$, while $y_i$ is an eigenvalue associated with
${\cal D}_2$.

All spectral correlation functions can be derived from the four kernels above. 
In the special case of a symmetric weight, $w^{(N_f)}(x,y)=w^{(N_f)}(y,x)$,
the polynomials coincide, $P_k^{(N_f)}=Q_k^{(N_f)}$, leading to coinciding transforms 
$\chi_k^{(N_f)}=\hchi_k^{(N_f)}$ and thus to a matching of the kernels 
$\hH_N^{(N_f)}(x,y)=H_N^{(N_f)}(y,x)$.

We call the correlation 
function of a set containing $n$ eigenvalues of ${\cal D}_1$ and $k$ eigenvalues of ${\cal D}_2$ 
an {\em $(n,k)$-correlation function}
\beqn
\label{Revrep}
R_{(n,k)}^{(N_f)}(\{x\}_n,\{y\}_k) &\equiv&
\frac{N!^2}{(N-n)!(N-k)!}\frac{1}{{\cal Z}_{\nu}^{(N_f)}(\{m_1\},\{m_2\})} \\
&\times&
\int_0^{\infty}  \prod_{i=n+1}^N dx_i \prod_{j=k+1}^N dy_j 
\det\left[w^{(N_f)}(x_i,y_j)\right]
\Delta_N(\{x^2\})\Delta_N(\{y^2\})
. \nn
\eeqn
Note for later comparison that these include both the connected and
disconnected parts.

Eynard and Mehta \cite{EM} have proven a generalized 
``Dyson Theorem'' for all correlation functions of random multi-matrix theories coupled
in a chain. Adapted to our chiral Two-Matrix Theory it can
be stated as follows:
\beq
R_{(n,k)}^{(N_f)}(\{x\}_n,\{y\}_k) ~=~ 
\det_{1\leq i_1,i_2\leq n;\ 1\leq  j_1,j_2\leq k}\left[
\begin{array}{cc}
H_N^{(N_f)}(x_{i_1},x_{i_2}) & M_N^{(N_f)}(x_{i_1},y_{j_2})-w^{(N_f)}(x_{i_1},y_{j_2})\\
K_N^{(N_f)}(y_{j_1},x_{i_2}) & \hH_N^{(N_f)}(y_{j_1},y_{j_2})\\
\end{array}
\right].
\label{allRnk}
\eeq
The simplest example, 
the spectral density itself equals
\beq
R_{(1,0)}^{(N_f)}(x) = H_N^{(N_f)}(x,x) ~,
\eeq
and similar for $R_{(0,1)}^{(N_f)}(y)$.
In the symmetric case (isospin) the densities are the same for the
 two operators ${\cal D}(\mu)$ and ${\cal D}(-\mu)$ as is obvious 
from the symmetry $\Psi \leftrightarrow -\Psi$ in the partition function
(\ref{ZNf}). 

More examples will be given in the next section.
If all eigenvalues of one kind are integrated out this reduces to the known
result for the one-matrix theory (up to a $\mu$ dependent rescaling),
as can be seen by generalizing the argument
from \cite{MS} for the unquenched non-chiral theory:
\beq
R_{(n,k=0)}^{(N_f)}(\{x\}_n) ~=~ \det\left[
  H_N^{(N_f)}(x_i,x_j)\right]\ ,
\label{Rn}
\eeq
and similarly for $R_{(n=0,k)}^{(N_f)}(\{y\}_k)$.


\sect{Explicit Computations}\label{sec-expcomp}

In this section we explicitly determine the biorthogonal polynomials
fulfilling the condition (\ref{bio}). It is instructive to consider
first the quenched case $N_f=0$ where there are no flavor determinants
in the integration measure. This is not only because matters simplify
then, but because the general solution can be expressed in terms of
quenched quantities.  Thus after having gone through the quenched case
in detail, we present the general solution for any $N_f$. We also
take the large-$N$ scaling limit, and thus make contact to the field
theory calculations of refs. \cite{DHSS,DHSST}.


\subsection{\sc The quenched case}

The analog of quenching in the Random Matrix Theory formulation is easily
achieved: We simply remove the explicit determinant factors in the measure. The
theory remains chiral, and we still perform the change of variables (\ref{Phipm}).
The weight function (\ref{weight}) then simplifies considerably and the
orthogonal polynomials can be readily obtained as shown in
 Appendices \ref{Lbiop} and \ref{wexp}.  The result is
\beqn
P_n(x^2) &=& n! (-N \tau c_1)^{-n} L_n^{(\nu)} (N \tau c_1 x^2) \nn\\
Q_n(y^2) &=& n! (-N \tau c_2)^{-n} L_n^{(\nu)} (N \tau c_2 y^2)
\label{polP}
\eeqn
with 
\beq
\tau = 1-d^2/(c_1 c_2)
\eeq 
from Appendix \ref{wexp} and
the normalization constant $h_k$ is given in (\ref{wexp-hn}).
Here and in the following we drop the superscript $(N_f=0)$ for the
quenched quantities.
The transforms (\ref{chidef}) can also be found from (\ref{wexp-bt}) along
with (\ref{wexp-wop}) and (\ref{wexp-nop})
\beqn
\chi_k(y)&=& \frac{k!}{2Nc_1(-N\tau c_2)^k}\left(\frac{d}{c_1}\right)^{2k+\nu}
y^{2\nu+1}e^{-N\tau c_2y^2}L_k^{(\nu)} (N \tau c_2 y^2) \cr
\hat{\chi}_k(x)&=& \frac{k!}{2Nc_2(-N\tau c_1)^k}\left(\frac{d}{c_2}\right)^{2k+\nu}
x^{2\nu+1}e^{-N\tau c_1x^2}L_k^{(\nu)} (N \tau c_1 x^2) \ . 
\eeqn
These are proportional to the polynomials in eq. (\ref{polP}).

Having the required functions, we can now write down the related kernels.
These can be obtained from (\ref{wexp-kern}) which gives
\beqn
K_N(x,y) &=&
 \frac{4 N (N \tau c_1 c_2)^{\nu+1}}{d^{\nu}}
 \sum_{k=0}^{N-1} \frac{k!}{(k+\nu)!} \left(\frac{c_1 c_2}{d^2}\right)^k
 L_k^{(\nu)}\left(N\tau c_1 x^2\right)
 L_k^{(\nu)}\left(N\tau c_2 y^2\right), \\
H_N(x_1,x_2) &=&
 2 (N \tau c_1)^{\nu+1} x_1^{2\nu+1} e^{-N\tau c_1 x_1^2}
 \sum_{k=0}^{N-1} \frac{k!}{(k+\nu)!}
 L_k^{(\nu)}\left(N\tau c_1 x_1^2\right)
 L_k^{(\nu)}\left(N\tau c_1 x_2^2\right), \cr
\hH_N(y_1,y_2) &=&
 2 (N \tau c_2)^{\nu+1} y_2^{2\nu+1} e^{-N\tau c_2 y_2^2}
 \sum_{k=0}^{N-1} \frac{k!}{(k+\nu)!}
 L_k^{(\nu)}\left(N\tau c_2 y_1^2\right)
 L_k^{(\nu)}\left(N\tau c_2 y_2^2\right), \cr
M_N(x,y) &=& 
 \tau (N \tau d)^\nu (xy)^{2\nu+1} e^{-N\tau (c_1 x^2 + c_2 y^2)}
 \sum_{k=0}^{N-1} \frac{k!}{(k+\nu)!} \left(\frac{d^2}{c_1 c_2}\right)^k
 L_k^{(\nu)}\left(N\tau c_1 x^2\right)
 L_k^{(\nu)}\left(N\tau c_2 y^2\right) ~. \nn
\label{kernelQ}
\eeqn

We now take the microscopic scaling limit where $N \to \infty$ and
$\hat{\mu}_f^2 = 2\mu_f^2 N$ and $\hat{m} = 2mN$ are kept fixed.
We define microscopic variables by 
\beqn
\hat{x} &\equiv& 2xN \cr 
\hat{y} &\equiv& 2yN \ .
\eeqn
Making use of the asymptotic behavior of Laguerre polynomials,
\beq
\lim_{n\to\infty}L_n^{(a)}(x^2) \sim n^a (nx^2)^{-a/2}J_a(2\sqrt{n}x) 
\label{Lasymp}
\eeq
where $J_k(x)$ is a Bessel function, we can turn sums into integrals
$1/N\sum_{k=0}^{N-1}\to \int_0^1 dt$ with $t=k/N$
\beq
\sum_{k=0}^{N-1} \frac{k!}{(k+\nu)!} \left(\frac{c_1 c_2}{d^2}\right)^{s k}
 L_k^{(\nu)}\left(N\tau c_x x^2\right) L_k^{(\nu)}\left(N\tau c_y y^2\right)
\sim
 \frac{1}{(\hx\hy)^\nu}\int_0^1 dt \,
e^{s\hat{\delta}^2 t/2}
 J_{\nu}(\hat{x}\sqrt{t})J_{\nu}(\hat{y}\sqrt{t}) ~.
\eeq
Here $s$ can have the values $0, \pm 1$ and $c_x$ and $c_y$ can be either
of $c_1$ or $c_2$.
We have also used
\beq
\left( \frac{c_1 c_2}{d^2} \right)^{s k}
= \left( \frac{(1+\mu_1^2)(1+\mu_2^2)}{(1+\mu_1\mu_2)^2} \right)^{s k}
\approx \left( 1 + \frac{(\hat\mu_1-\hat\mu_2)^2}{2N} \right)^{s k}
\to \exp\left( \frac{s}{2} \, \hat\delta^2 t \right)
\label{mut}
\eeq
with $\hat\delta = \hat\mu_2 - \hat\mu_1$. In this scaling limit the kernels
therefore take the form  
\beqn
\lim_{N\to\infty}
 K_N\left(x=\frac{\hx}{2N},y=\frac{\hy}{2N}\right) &=& 
 2^{2\nu+4}N^{2\nu+4} (\hx\hy)^{-\nu} \hd^{-2}
\mathcal{I}^+(\hx,\hy)
\label{Ks}\\
&& \cr
\lim_{N\to\infty}
H_N\left(x=\frac{\hx}{2N},y=\frac{\hy}{2N}\right)&=& 
2 N \hx^{\nu+1}\hy^{-\nu}
\mathcal{I}^0(\hx,\hy)
\label{Hs} \\
&=& \lim_{N\to\infty}
 \hH_N\left(y=\frac{\hy}{2N},x=\frac{\hx}{2N}\right)
\cr
&& \cr
\lim_{N\to\infty}
 M_N\left(x=\frac{\hx}{2N},y=\frac{\hy}{2N}\right)
&=& 
2^{-2\nu-2}N^{-2\nu-2} (\hx\hy)^{\nu+1}\hd^2
\mathcal{I}^-(\hx,\hy)
\label{Ms}
\eeqn
where we introduced the abbreviated notation
\beqn
{\cal I}^0(\hat{x},\hat{y}) &\equiv&
 \frac12 \int_0^1 dt\, J_{\nu}(\hat{x}\sqrt{t})J_{\nu}(\hat{y}\sqrt{t})
=
 \frac{\hat{x}J_{\nu+1}(\hat{x})J_{\nu}(\hat{y}) - \hat{y}J_{\nu+1}(\hat{y})
J_{\nu}(\hat{x})}{\hat{x}^2-\hat{y}^2} \cr
{\cal I}^\pm(\hat{x},\hat{y}) &\equiv& \frac12
 \int_0^1 dt \, e^{\pm\hd^2 t/2}J_{\nu}(\hat{x}\sqrt{t})J_{\nu}(\hat{y}\sqrt{t}) ~.
\label{shortint}
\eeqn
The weight becomes
\beqn
\hat{w}^{(0)}(\hx,\hy)
 = \lim_{N\to\infty} w^{(0)}(\frac{\hx}{2N},\frac{\hy}{2N})
 = \left(\frac{\hx\hy}{4N^2}\right)^{\nu+1}
   \exp\left(-\frac{\hx^2+\hy^2}{2\hat{\delta}^2}\right)
   I_{\nu}\left(\frac{\hat{x}\hat{y}}{\hat{\delta}^2}\right) ~,
\eeqn
using the abbreviation
\beq
\tilde{\mathcal{I}}^-(\hx,\hy) ~\equiv~
\frac{1}{\hd^2} \exp\left(-\frac{\hx^2+\hy^2}{2\hd^2}\right)
 I_{\nu}\left(\frac{\hx\hy}{\hd^2}\right) 
-  \mathcal{I}^-(\hx,\hy) ~.
\eeq
Notice that the $\mu$'s only appear through their difference.
Despite the relation (\ref{w-M}) 
between the weight and the kernel $M$ we will derive in appendix \ref{wexp}
our large-$N$ results of the two {\it differ}. The reason for this is that the
large-$N$ limit is not uniform: while we have rescaled the chemical potentials
with $N$, also called ``weak non-Hermiticity'' in the context of real $\mu$, 
another limit exists where the chemical potentials
are not scaled, including a different rescaling of the eigenvalues with $N$. In
this limit called ``strong non-Hermiticity'' for real $\mu$, which can also be obtained by
letting $\delta\to\infty$, the large-$N$ results of weight and $M$ will match,
as suggested by relation (\ref{w-M}). For more details on this limit we
refer to \cite{G1}. 

The correlation functions in the microscopic scaling limit are defined as
follows for general $N_f$:
\beq
\rho_{(n,k)}^{(N_f)}(\{\hat{x}\}_n,\{\hat{y}\}_k)~\equiv~ \lim_{N\to\infty}\frac{1}{(2N)^{n+k}}
R_{(n,k)}^{(N_f)}\left(\{x=\frac{\hx}{2N}\}_n,\{y=\frac{\hy}{2N}\}_k\right) 
\label{microdef}
\eeq
Inserting eqs. (\ref{Ks}) - (\ref{Ms}) into the definition we arrive at the
following result
\beq
\rho_{(n,k)}(\{\hat{x}\}_n,\{\hat{y}\}_k) =
 \prod_{i=1}^n \hx_i\prod_{j=1}^k \hy_j
\times\det_{1\leq i_1,i_2\leq n;\ 1\leq  j_1,j_2\leq k}\left[
\begin{array}{cc}
\mathcal{I}^0(\hx_{i_1},\hx_{i_2})
&
-\tilde{\mathcal{I}}^-(\hx_{i_1},\hy_{j_2})
\\
\mathcal{I}^+(\hy_{j_1},\hx_{i_2})
&
\mathcal{I}^0(\hy_{j_1},\hy_{j_2})
\end{array}
\right]
\label{allRnkS}
\eeq
after taking out common factors.

As a special case we note that the (1,0)-correlation function reduces
to the known Bessel kernel taken at coincident points, yielding the
$\hmu_f$-independent
\beq
\rho_{(1,0)}(\hat{x}) ~=~ \frac{\hat{x}}{2}\left[J_{\nu}^2(\hat{x})
- J_{\nu+1}(\hat{x})J_{\nu-1}(\hat{x})\right] ~, \label{rhoQ}
\eeq 
which is the known quenched expression from the one-matrix theory
\cite{VZ}.  The $\hmu_f$-independence of this result is particularly easy
to understand from the effective Lagrangian, where the corresponding
$B$-matrix is proportional to the unit matrix, and thus can be shifted
away.  Similarly, the (2,0)-correlation function of this quenched
theory becomes
\beq
\rho_{(2,0)}(\hat{x}_1,\hat{x}_2) =
\rho_{(1,0)}(\hat{x}_1)\rho_{(1,0)}(\hx_2)\ -\ 
\hx_1\hx_2\left[
\frac{\hat{x}_1J_{\nu+1}(\hat{x}_1)J_{\nu}(\hat{x}_2) - \hat{x}_2J_{\nu+1}(\hat{x}_2)
J_{\nu}(\hat{x}_1)}{\hat{x}_1^2-\hat{x}_2^2}\right]^2
\label{rhoQ20}
\eeq
which again is $\hmu_f$-independent, and agrees with the standard
expression for the $\mu_f=0$ theory \cite{VZ}. One easily sees how
this generalizes to the the general $(n,0)$-correlator in the quenched
theory: the usual $n$-point correlation function of the one-matrix
theory \cite{ADMN,DN} is recovered from eq. (\ref{allRnk}), taking out
and canceling the prefactors from eq. (\ref{Hs}) from the
determinant.

A particularly important quantity is the mixed (1,1)-correlation function 
$\rho_{(1,1)}(\hat{x},\hat{y})$ which reads
\beq
\rho_{(1,1)}(\hat{x},\hat{y}) 
~\equiv~ \rho_{(1,0)}(\hat{x})\rho_{(0,1)}(\hy)\ +\
\rho_{(1,1)}^{conn}(\hat{x},\hat{y}) \ . 
\label{rhoQ11}
\eeq
The connected part is given by
\beqn
\rho_{(1,1)}^{conn}(\hat{x},\hat{y}) &=&
\hat{x}\hat{y} ~ \mathcal{I}^+(\hy,\hx) ~
\tilde{\mathcal{I}}^-(\hx,\hy) ~.
\label{rhoQ11int}
\eeqn
After making the required substitution
$\hd^2 \to 4 VF_{\pi}^2\mu^2$ and the identification
$\hat{x} = \lambda_+\Sigma V$, $\hat{y} = \lambda_-\Sigma V$,
this agrees exactly with what was previously computed from the
effective low-energy QCD Lagrangian \cite{DHSS}. In that framework,
however, the general correlation functions $\rho_{(n,k)}^{conn}$ have
not been obtained.

Another interesting case is the quenched limit when $\mu$ is not scaled
microscopically, as discussed above. In this limit the upper right corner of
the determinant in eq.  
(\ref{allRnkS}) vanished due to the matching of the weight and large-$N$ kernel
$M$. This leads to a factorisation of all correlation functions into a
product of determinants over $H$ and $\hH$.


\subsection{\sc Two and more flavors}\label{2plus}

To find the biorthogonal polynomials it is convenient to use the following
general expression in the eigenvalue representation
\beqn
P_k^{(N_f)}(x^2) &=& (-)^k (ix)^{-\nu}
 \langle \det[ \mathcal{D}_1 +ix ]\rangle_{{\cal Z}_{\nu}^{(N_f)}} \cr
&=& \frac{(-)^k{\cal N}(\{m_{1,2}\})}{{\cal Z}_{\nu}^{(N_f)}}
 \int_0^{\infty} \prod_{i=1}^k\left(dx_i dy_i (x_i y_i)^{\nu+1}(x_i^2 - x^2)
\prod_{f1=1}^{N_1} (x_i^2+m_{f1}^2)
\prod_{f2=1}^{N_2} (y_i^2+m_{f2}^2)\right) \cr
&\times&\Delta_k(\{x^2\})\Delta_k(\{y^2\})
\det\left[I_{\nu}(2 d N x_i y_j)\right] 
e^{-N \sum_i c_1 x_i^2 + c_2 y_i^2 }
\label{charP}
\eeqn
generalising the observation made in the appendix of \cite{E}. As also noted
there a similar relation holds for 
\beq
Q_k^{(N_f)}(y^2) ~=~ (-)^k (iy)^{-\nu} \langle
\det \mathcal{D}_2 +iy \rangle_{{\cal Z}_{\nu}^{(N_f)}}\ .
\label{charQ}
\eeq
The expectation values are taken
with respect to the partition function in eq. (\ref{ZNfpm}) here\footnote{The
  partition function used here contains $k$ eigenvalues, whereas the 
  explicit factor of $N$ in the exponent remains fixed.}.
The prefactor $(-)^k (ix)^{-\nu}$ ensures that $P_k^{(N_f)}(x^2)$ is a monic
polynomial of order $k$, and similarly for $Q_k^{(N_f)}(y^2)$. 
Eq. (\ref{charP}) (eq. (\ref{charQ})) is recognized as being a partition 
function of one additional $\mathcal{D}_1$-flavor ($\mathcal{D}_2$-flavor) 
with purely imaginary mass $ix$ ($iy$),
\beqn
P_k^{(N_f)}(x^2) 
&=& (-)^k (ix)^{-\nu}
\frac{{\cal Z}_{\nu}^{(N_1+1,N_2)}(\{m_1\},ix,\{m_2\})}{{\cal Z}_{\nu}^{(N_f)}(\{m_1\},\{m_2\})} ~, \cr
Q_k^{(N_f)}(y^2) 
&=& (-)^k (iy)^{-\nu}
\frac{{\cal Z}_{\nu}^{(N_1,N_2+1)}(\{m_1\},\{m_2\},iy)}{{\cal Z}_{\nu}^{(N_f)}(\{m_1\},\{m_2\})}
\label{ZP}
\eeqn
indicated explicitly in the superscript.
In the large-$N$ scaling limit this gives a particularly useful 
representation \cite{D} since
compact analytical expressions
can be obtained from refs. \cite{SplitV,AFV}, where the involved finite-volume partition functions 
were derived in the effective field theory framework.
Alternatively, we can re-derive those expressions from our two-matrix theory by
substituting the corresponding large-$N$ expressions from the previous subsection into eq. (\ref{ZP}),
and making use of the general expression for partition functions given in appendix \ref{allZ}.

Instead of inserting the relation eq. (\ref{ZP}) into the definition of the
kernel eq. (\ref{K}) a far more useful relation exists, giving the kernel
directly in terms of a single partition function:
\beqn
K_{N+1}^{(N_f)}(y,x) &=& \frac{(-xy)^{-\nu}}{h_N^{(N_f)}} \langle \det[ \mathcal{D}_1 +i x ]
\det[ \mathcal{D}_2 +i y ]\rangle_{{\cal
    Z}_{\nu}^{(N_f)}}
\label{charK}
\eeqn
This relation follows in complete analogy to \cite{AV} after going through the
same steps as in appendix \ref{allZ}. 
The highest polynomials in  $K_{N+1}^{(N_f)}(y,x)$ are of degree $N$, and in
order to match we have to divide by their coefficient ${h_N^{(N_f)}}$, given
by \cite{AV,B}
\beq
h_N^{(N_f)}~=~(-)^{N_1+N_2}h_{N+N_2}
\frac{\det\left[ K_{N+N_2+1}(im_{l2},im_{k1}) \ P_{N+N_2+1}((im_{k1})^2)\ldots
    P_{N+N_1}((im_{k1})^2)\right]}{\det\left[ K_{N+N_2}(im_{l2},im_{k1}) \
    P_{N+N_2}((im_{k1})^2)\ldots  P_{N+N_1-1}((im_{k1})^2)\right]} 
\label{hNf}
\eeq
for $N_1\geq N_2$ (for $N_2 > N_1$ simply exchange the $1$ with the $2$
quantities as in appendix \ref{allZ}). In the large-$N$ limit the two
determinants will cancel, leaving only the quenched norm.
In terms of partition functions with two additional
flavors this can be written as
\beq
 K_{N+1}^{(N_f)}(y,x)= (-xy)^{-\nu}\frac{{\cal
 Z}_{\nu}^{(N_1+1,N_2+1)}(\{m_1\},ix,\{m_2\},iy)}{h_N^{(N_f)}{\cal
     Z}_{\nu}^{(N_f)}(\{m_1\},\{m_2\})} ~.   
\label{ZK}
\eeq
Compared to the polynomial we have added a second flavor with 
chemical potential  $\mu_2$, related to the {\it first} argument 
of the kernel. 
If on the contrary we had added a second flavor of the same kind, $\langle \det[ \mathcal{D}_1 +i x_1 ]
\det[ \mathcal{D}_1 +ix_2 ]\rangle_{{\cal Z}_{\nu}^{(N_f)}}$, we would
have obtained a determinant of polynomials instead of the kernel (see appendix
\ref{allZ}). Only when $\mu_1=\mu_2=0$ the two are related through the 
Christoffel-Darboux identity.

The functions $\chi_k^{(N_f)}(y)$ and $\hchi_k^{(N_f)}(x)$ can then be computed
from their definition (\ref{chidef})
\beqn
\chi_k^{(N_f)}(y) = \int_0^\infty dx  
w^{(N_f)}(x,y)(-)^k 
(ix)^{-\nu}\frac{{\cal Z}_{\nu}^{(N_1+1,N_2)}(\{m_1\},ix,\{m_2\})}{{\cal
    Z}_{\nu}^{(N_f)}(\{m_1\},\{m_2\})} \cr
\hchi_k^{(N_f)}(x) = \int_0^\infty dy  
w^{(N_f)}(x,y)(-)^k 
(iy)^{-\nu}\frac{{\cal Z}_{\nu}^{(N_1,N_2+1)}(\{m_1\},\{m_2\},iy)}{{\cal
    Z}_{\nu}^{(N_f)}(\{m_1\},\{m_2\})} \ .
\label{Zchi}
\eeqn
Similarly the remaining kernels can be obtained from
eq. (\ref{ZK}) by transforming one or both arguments
\beqn
H_{N+1}^{(N_f)}(x_1,x_2) &=&
\int_0^\infty dy w^{(N_f)}(x_1,y)
(-x_2y)^{-\nu}\frac{{\cal
 Z}_{\nu}^{(N_1+1,N_2+1)}(\{m_1\},ix_2,\{m_2\},iy)}{h_N^{(N_f)}{\cal Z}_{\nu}^{(N_f)}(\{m_1\},\{m_2\})}
\label{ZH}
\\
\hH_{N+1}^{(N_f)}(y_1,y_2) &=&
\int_0^\infty dx w^{(N_f)}(x,y_2)
(-xy_1)^{-\nu}\frac{{\cal
 Z}_{\nu}^{(N_1+1,N_2+1)}(\{m_1\},ix,\{m_2\},iy_1)}{h_N^{(N_f)}{\cal Z}_{\nu}^{(N_f)}(\{m_1\},\{m_2\})}
\label{ZHh}
\\
M_{N+1}^{(N_f)}(x,y) &=& 
\int_0^\infty dudv w^{(N_f)}(x,u) w^{(N_f)}(v,y) 
(-vu)^{-\nu}\frac{{\cal
 Z}_{\nu}^{(N_1+1,N_2+1)}(\{m_1\},iv,\{m_2\},iu)}{h_N^{(N_f)}{\cal
    Z}_{\nu}^{(N_f)}(\{m_1\},\{m_2\})} \ .\cr
&&
~ \label{ZM}
\eeqn
Inserting them into eq. (\ref{allRnk}) this provides the general
$(n,k)$-correlation function at finite-$N$  
of the theory with arbitrary $N_f$.
Since they are expressed in terms of partition functions the microscopic
scaling limit can easily be taken, as we have mentioned above.

\begin{figure*}[ht]
  \unitlength1.0cm
  \epsfig{file=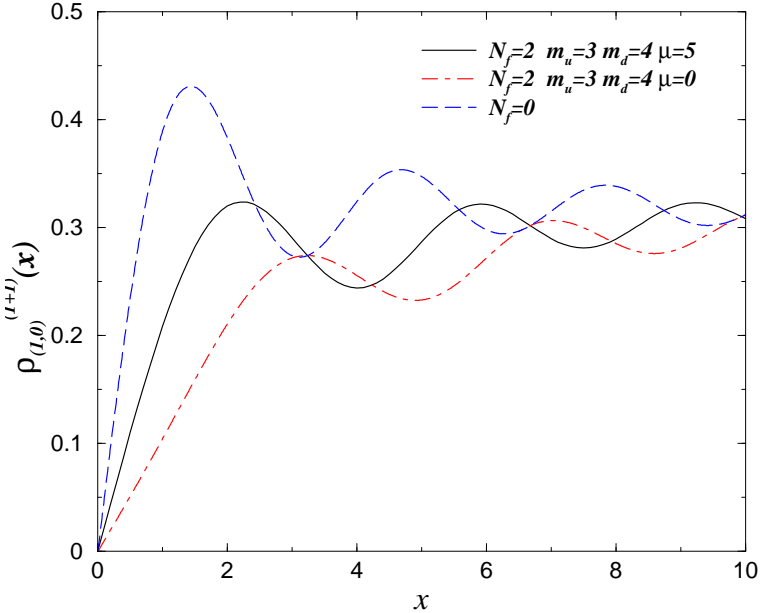,clip=,width=10cm}
  \caption{ 
    \label{fig:1} The eigenvalue density for $N_f=2$ with imaginary isospin
    chemical potential $\hat{\mu}=\hd/2=0$ and $5$ and quark masses
    $\hat{m}_u=3$, $\hat{m}_d=4$. 
The quenched density eq. (\ref{rhoQ}) of the one-matrix theory ($N_f=0$) is
given for comparison.
}
\end{figure*}

We now specialize to the physically most interesting case of two flavors
 $N_1=N_2=1$, with quark masses $m_u$ and $m_d$, respectively.
We first focus on the (1,1)-correlation function
 $\rho_{(1,1)}^{(1+1)\,conn}(\hat{x},\hat{y})$ in order to
compare with the known result \cite{DHSST}. 
The first building block  we need, the unquenched kernel, is given by 
\beq 
K_{N+1}^{(1+1)}(y,x)=\frac{h_{N+1}}{h_N^{(N_f)}(x^2+m_u^2)(y^2+m_d^2)K_{N+1}(im_d,im_u)}
\det\left[
\begin{array}{cc}
K_{N+1}(im_d,im_u) & K_{N+1}(im_d,x)\\
K_{N+1}(y,im_u) & K_{N+1}(y,x)\\
\end{array}
\right] , 
\label{K11}
\eeq
where we used that the quenched kernel is even in its arguments.
The second building block is obtained from that by integrating twice from eq. (\ref{M})
\beqn
M_{N+1}^{(1+1)}(x,y)&=&\int_0^\infty dudv\ w^{(1+1)}(x,u)w^{(1+1)}(v,y)
K_{N+1}^{(1+1)}(u,v)\cr
&=&\frac{h_{N+1}(x^2+m_u^2)(y^2+m_d^2)}{h_N^{(N_f)}K_{N+1}(im_d,im_u)}
\det\left[
\begin{array}{cc}
K_{N+1}(im_d,im_u) & \hH_{N+1}(im_d,y)\\
H_{N+1}(x,im_u) & M_{N+1}(x,y)\\
\end{array}
\right] . 
\label{M11}
\eeqn
Taking the integrals inside the determinant leads to the appearance of all
different quenched  kernels. In order to take the large-$N$ limit we can thus
use the quenched results eqs. (\ref{Ks}) - (\ref{Ms}).
We therefore obtain for the two microscopic building blocks 
\beqn 
\lim_{N\to\infty}K_{N+1}^{(1+1)}(y,x)
&=& \frac{(2N)^{2\nu+8}(\hx\hy)^{-\nu}\hd^{-2}}
 {(\hx^2+\hm_u^2)(\hy^2+\hm_d^2)}
\left( \mathcal{I}^+(\hy,\hx) -
 \frac{\mathcal{I}^+(\hy,i\hat{m}_u) \mathcal{I}^+(i\hat{m}_d,\hx)}
 {\mathcal{I}^+(i\hat{m}_d,i\hat{m}_u)}
\right)
\label{K11s}
\eeqn
and
\beqn
\lim_{N\to\infty}M_{N+1}^{(1+1)}(x,y)
&=& \frac{\hd^2 (\hx\hy)^{\nu+1} (\hx^2+\hm_u^2)(\hy^2+\hm_d^2)}
{(2N)^{2\nu+6}}
\left( \mathcal{I}^-(\hx,\hy) -
 \frac{\mathcal{I}^0(\hx,i\hat{m}_u) \mathcal{I}^0(i\hat{m}_d,\hy)}
 {\mathcal{I}^+(i\hat{m}_u,i\hat{m}_d)}
\right) ~
\label{M11s}
\eeqn
along with the weight
\beq
\lim_{N\to\infty} w^{(1+1)}(x,y)~=~
\frac{(\hx^2+\hm_u^2)(\hy^2+\hm_d^2)}{(2N)^{4}}\ 
\frac{(\hx\hy)^{\nu+1}}{(2N)^{2\nu+2}}
I_\nu\left(\frac{\hx\hy}{\hd^2}\right)e^{-\frac{\hx^2+\hy^2}{2\hd^2}} ~.
\eeq
Here the norms have canceled in the large-$N$ limit, and the explicit mass
factors will also drop out when multiplying all together in the next
step. Using eqs. (\ref{allRnk}) and (\ref{microdef}) we first consider the
following connected part of the two point function:
\beqn\label{rho2conn}
\rho_{(1,1)}^{(1+1)\,conn}(\hx,\hy) &\equiv& \lim_{N\to\infty}
\frac{1}{(2N)^2}K_{N+1}^{(1+1)}(y,x)
\left(w^{(1+1)}(x,y) -M_{N+1}^{(1+1)}(x,y)\right) \\
&=& \hx\hy
 \left( \mathcal{I}^+(\hy,\hx) -
  \frac{\mathcal{I}^+(\hy,i\hat{m}_u) \mathcal{I}^+(i\hat{m}_d,\hx)}
  {\mathcal{I}^+(i\hat{m}_d,i\hat{m}_u)} \right)
 \left( \tilde{\mathcal{I}}^-(\hx,\hy) +
  \frac{\mathcal{I}^0(\hx,i\hat{m}_u) \mathcal{I}^0(i\hat{m}_d,\hy)}
  {\mathcal{I}^+(i\hat{m}_u,i\hat{m}_d)} \right) ~. \nn
\eeqn
Substituting $\hd^2 \to 4 VF_{\pi}^2\mu^2$ and 
$\hat{x} = \lambda_+\Sigma V, \hat{y} = \lambda_-\Sigma V$,
this again agrees with the unquenched result 
from effective low-energy QCD \cite{DHSST}.
For completeness we also give the other two building blocks for this
theory. Following the definition (\ref{H}) we have 
\beqn
\label{H11}
H_{N+1}^{(1+1)}(x_1,x_2)&=&\int_0^\infty dy\ w^{(1+1)}(x_1,y)
K_{N+1}^{(1+1)}(y,x_2)\\
&=&\frac{h_{N+1}(x_1^2+m_u^2)}{h_N^{(N_f)}(x_2^2+m_u^2)K_{N+1}(im_d,im_u)}
\det\left[\!
\begin{array}{cc}
K_{N+1}(im_d,im_u) & K_{N+1}(im_d,x_2)\\
H_{N+1}(x_1,im_u) & H_{N+1}(x_1,x_2)\\
\end{array}\! 
\right] \nn
\eeqn
leading to 
\beqn
\lim_{N\to\infty}H_{N+1}^{(1+1)}(x_1,x_2)
&=& \frac{2N \hx_1^{\nu+1} (\hx_1^2+\hm_u^2)}{\hx_2^{\nu} (\hx_2^2+\hm_u^2)}
\left( \mathcal{I}^0(\hx_1,\hx_2) -
 \frac{\mathcal{I}^0(\hx_1,i\hat{m}_u) \mathcal{I}^+(i\hat{m}_d,\hx_2)}
 {\mathcal{I}^+(i\hat{m}_d,i\hat{m}_u)} \right) ~.
\label{H11s}
\eeqn
For the last block we obtain from eq. (\ref{Hh})  
\beqn
\label{Hh11}
\hH_{N+1}^{(1+1)}(y_1,y_2)&=&\int_0^\infty dx\ w^{(1+1)}(x,y_2)
K_{N+1}^{(1+1)}(y_1,x)\\
&=&\frac{h_{N+1}(y_2^2+m_d^2)}{h_N^{(N_f)}(y_1^2+m_d^2)K_{N+1}(im_d,im_u)}
\det\left[\!
\begin{array}{cc}
K_{N+1}(im_d,im_u) & \hH_{N+1}(im_d,y_2)\\
K_{N+1}(y_1,im_u) & \hH_{N+1}(y_1,y_2)\\
\end{array}
\!\right] \nn
\eeqn
leading to 
\beqn
\lim_{N\to\infty}\hH_{N+1}^{(1+1)}(y_1,y_2)
&=& \frac{2N \hy_2^{\nu+1} (\hy_2^2+\hm_d^2)}{\hy_1^{\nu} (\hy_1^2+\hm_d^2)}
\left( \mathcal{I}^0(\hy_1,\hy_2) -
 \frac{\mathcal{I}^+(\hy_1,i\hat{m}_u) \mathcal{I}^0(i\hat{m}_d,\hy_2)}
 {\mathcal{I}^+(i\hat{m}_d,i\hat{m}_u)} \right) ~.
\label{Hh11s}
\eeqn
As a new microscopic result we have the spectral densities 
\beqn
\rho_{(1,0)}^{(1+1)}(\hx) &=& \lim_{N\to\infty} \frac{1}{2N} H_{N+1}^{(1+1)}(x,x)\cr
\rho_{(0,1)}^{(1+1)}(\hy) &=& \lim_{N\to\infty} \frac{1}{2N} \hH_{N+1}^{(1+1)}(y,y)\ ,
\eeqn
given by eqs. (\ref{H11s}) and (\ref{Hh11s}) at equal arguments divided by $2N$
respectively. Note that the two functions coincide. At $\hmu_1=\hmu_2=0$ they
 reproduce the known result \cite{DN} 
for two different flavors. For a plot of the unquenched eigenvalue density see figure \ref{fig:1}.

To finish this subsection we give the result for the most general correlation
function in the two flavor case. If follows by inserting the above building
blocks into eq. (\ref{allRnk}) and taking out common factors:
\beqn
\label{allRnk11S}
\rho_{(n,k)}^{(1+1)}(\{\hat{x}\}_n,\{\hat{y}\}_k) &=&
 \prod_{i=1}^n \hx_i\prod_{j=1}^k
 \hy_j \left({\cal I}^+(i\hm_u,i\hm_d)\right)^{-n-k} \\
&\times& \det_{\stackrel{1\leq i_1,i_2\leq n}{\ 1\leq  j_1,j_2\leq k}}
\left[
\begin{array}{cc}
 {\cal I}^+(i\hm_u,i\hm_d)    {\cal I}^0(\hx_{i_1},\hx_{i_2}) &
-{\cal I}^+(i\hm_u,i\hm_d)    \tilde{{\cal I}}^-(\hy_{j_1},\hx_{i_2}) \\
-{\cal I}^0(\hx_{i_1},i\hm_u) {\cal I}^+(\hx_{i_2},i\hm_d) &
-{\cal I}^0(\hy_{j_1},i\hm_d) {\cal I}^0(\hx_{i_2},i\hm_u) \\
\\
 {\cal I}^+(i\hm_u,i\hm_d)    {\cal I}^+(\hy_{j_1},\hx_{i_2}) &
 {\cal I}^+(i\hm_u,i\hm_d)    {\cal I}^0(\hy_{j_1},\hy_{j_2}) \\
-{\cal I}^+(\hy_{j_1},i\hm_u) {\cal I}^+(\hx_{i_2},i\hm_d) &
-{\cal I}^+(\hy_{j_1},i\hm_u) {\cal I}^0(\hy_{j_2},i\hm_d)
\end{array}
\right]~. \nn
\eeqn


\subsection{\sc Partial quenching}

Here we consider a variant of the general $N_f$-theory of practical importance.
We introduce $N_f$ fermions with Dirac operator $D_1$ and set
$\mu_1=0$ then seek the spectral correlation functions between eigenvalues of
$D_1$ and $D_2$ for $\mu_2\ne0$. In lattice gauge theory
simulations this is a simple modification. The ensemble averages are
then taken with respect to a theory which is not coupled to a chemical
potential, but the observables (expressed in terms of ``valence
quarks'') are.  In the Random Matrix Theory formulation it can also be
introduced straightforwardly.

The corresponding Random Matrix Theory is based on the partition
function eq. (\ref{ZNfpm}) with $N_1=N_f$ and $N_2=0$ and we ask for
the mixed correlation function $R_{(1,1)}^{(N_f)}$ between two valence
quarks associated with matrices ${\cal D}_{1,2}$. For simplicity we
here restrict ourselves to the physically most interesting case of
$N_f=2$.  Here we have {\it two} sea quarks with masses $m_u$ and
$m_d$ and zero chemical potential $\mu_1=0$, while the valence quarks
have $\mu_1=0$ and $\mu_2=\delta$.

In the previous subsection the results for the most general $N_f=N_1+N_2$
flavors content have been given already for the polynomials in
eq. (\ref{ZP}) and the kernel in eq. (\ref{ZK}) from which we can also
compute the transforms (\ref{chidef}), (\ref{H}), (\ref{Hh}) and (\ref{M}). 
We just have to specify our case $N_1=2$, $N_2=0$ when applying eq.
(\ref{Zgenresult}).  We will first focus on
$\rho_{(1,1)}^{(2)\,conn}(\hx,\hy)$ in order to pinpoint the
difference between the $1+1$ flavor case eq. (\ref{rho2conn}) from the
previous subsection, which was also computed in an effective field
theory approach \cite{DHSST}. Next, we give the general
$(n,k)$-correlation function for this particular flavor combination.

The first building block for that from eq. (\ref{ZK}) is given by 
\beq
K_{N+1}^{(2)}(y,x)=\frac{h_{N}}{h_N^{(N_f)}(x^2+m_u^2)(x^2+m_d^2)}
\frac{\det\left[
\begin{array}{ccc}
K_{N+1}(y,im_u) & P_{N+1}(-m_u^2) & P_{N+2}(-m_u^2)\\
K_{N+1}(y,im_d) & P_{N+1}(-m_d^2) & P_{N+2}(-m_d^2)\\
K_{N+1}(y,x)    & P_{N+1}(x^2)    & P_{N+2}(x^2)\\
\end{array}
\right]}{\det\left[
\begin{array}{cc}
P_{N}(-m_u^2) & P_{N+1}(-m_u^2)\\
P_{N}(-m_d^2) & P_{N+1}(-m_d^2)\\
\end{array}
\right]}~.~
\label{K2}
\eeq
We observe a different structure of determinants compared to the corresponding
previous result eq. (\ref{K11}), marking the difference of partial quenching.
The microscopic large-$N$ limit is now easily taken using the formulas from the
previous subsection. The only difference is that we have to Taylor expand the
last columns in numerator and denominator as they become degenerate at
large-$N$. The expansion in $1/N$ maps into a Taylor expansion with respect to
the arguments and we obtain
\beq
\lim_{N\to\infty}K_{N+1}^{(2)}(y,x)
=\frac{(2N)^{2\nu+8}(\hx\hy)^{-\nu}\hd^{-2}}{(\hx^2+\hm_u^2)(\hx^2+\hm_d^2)}
\frac{\det\left[
\begin{array}{ccc}
 \mathcal{I}^+(\hy,i\hm_u) &
 J_\nu(i\hm_u) &
 i\hm_u J_{\nu+1}(i\hm_u) \\
 \mathcal{I}^+(\hy,i\hm_d) &
 J_\nu(i\hm_d) &
 i\hm_d J_{\nu+1}(i\hm_d) \\
 \mathcal{I}^+(\hy,\hx) &
 J_\nu(\hx) &
 \hx J_{\nu+1}(\hx)
\end{array}
\right]}{\det\left[
\begin{array}{cc}
J_\nu(i\hm_u) & i\hm_u J_{\nu+1}(i\hm_u)\\
J_\nu(i\hm_d) & i\hm_d J_{\nu+1}(i\hm_d)
\end{array}
\right]}~.
\label{K2s}
\eeq
The result for the second building block of our correlation function is 
obtained by integrating eq. (\ref{K2}) using the definition eq. (\ref{M})
\beq
M_{N+1}^{(2)}(x,y)=\frac{h_{N}}{h_N^{(N_f)}}(x^2+m_u^2)(x^2+m_d^2)
\frac{\det\left[
\begin{array}{ccc}
H_{N+1}(x,im_u) & P_{N+1}(-m_u^2) & P_{N+2}(-m_u^2)\\
H_{N+1}(x,im_d) & P_{N+1}(-m_d^2) & P_{N+2}(-m_d^2)\\
M_{N+1}(x,y)    & \chi_{N+1}(y)    & \chi_{N+2}(y)
\end{array}
\right]}{\det\left[
\begin{array}{cc}
P_{N}(-m_u^2) & P_{N+1}(-m_u^2)\\
P_{N}(-m_d^2) & P_{N+1}(-m_d^2)
\end{array}
\right]}~.
\label{M2}
\eeq
In addition to the asymptotics already computed we also now need one
of the transforms $\chi_N$. Being proportional to the polynomials,
eq. (\ref{wexp-bt}), this is an easy task, and we obtain an explicit
exponential $\hd$ dependence from the prefactors. We thus arrive at
\beq
\lim_{N\to\infty}M_{N+1}^{(2)}(x,y) =
\frac{(\hx\hy)^{\nu+1}(\hx^2+\hm_u^2)(\hx^2+\hm_d^2)}{(2N)^{2\nu+6}\hd^{2}}
\frac{\det\left[
\begin{array}{ccc}
 \mathcal{I}^0(\hx,i\hm_u) &
 J_\nu(i\hm_u) &
 i\hm_u J_{\nu+1}(i\hm_u) \\
 \mathcal{I}^0(\hx,i\hm_d) &
 J_\nu(i\hm_d) &
 i\hm_d J_{\nu+1}(i\hm_d) \\
 \mathcal{I}^-(\hx,\hy) &
 e^{-\hd^2/2}J_\nu(\hy) &
 e^{-\hd^2/2}G_\nu(\hy,\hd)
\end{array}
\right]}{\det\left[
\begin{array}{cc}
J_\nu(i\hm_u) & i\hm_u J_{\nu+1}(i\hm_u)\\
J_\nu(i\hm_d) & i\hm_d J_{\nu+1}(i\hm_d)
\end{array}
\right]} ~,
\label{M2s}
\eeq
where we have introduced the notation
\beq
G_\nu(\hy,\hd)=\hy J_{\nu+1}(\hy)+\hd^2J_{\nu}(\hy)~.
\label{G}
\eeq
Together with the asymptotic of the weight function, 
\beq
\lim_{N\to\infty} w^{(2)}(x,y)~=~
\frac{(\hx^2+\hm_u^2)(\hx^2+\hm_d^2)}{(2N)^{4}}\ 
\frac{(\hx\hy)^{\nu+1}}{(2N)^{2\nu+2}}
I_\nu\left(\frac{\hx\hy}{\hd^2}\right)e^{-\frac{\hx^2+\hy^2}{2\hd^2}} 
\eeq
we obtain the partially quenched two point function
\beq
\rho_{(1,1)}^{(2)\,conn}(\hx,\hy) =
\hx\hy \ \frac{
\left|\begin{array}{ccc}
 \mathcal{I}^+(\hy,i\hm_u) & J_\nu(i\hm_u) & i\hm_u J_{\nu+1}(i\hm_u) \\
 \mathcal{I}^+(\hy,i\hm_d) & J_\nu(i\hm_d) & i\hm_d J_{\nu+1}(i\hm_d) \\
 \mathcal{I}^+(\hy,\hx)    & J_\nu(\hx)    &  \hx   J_{\nu+1}(\hx)
\end{array}\right|
\left|\begin{array}{ccc}
 -\mathcal{I}^0(\hx,i\hm_u) & J_\nu(i\hm_u) & i\hm_u J_{\nu+1}(i\hm_u) \\
 -\mathcal{I}^0(\hx,i\hm_d) & J_\nu(i\hm_d) & i\hm_d J_{\nu+1}(i\hm_d) \\
 \tilde{\mathcal{I}}^-(\hx,\hy) & e^{-\hd^2/2}J_\nu(\hy) &
 e^{-\hd^2/2}G_\nu(\hy,\hd)
\end{array}\right|
}{
\left|\begin{array}{cc}
 J_\nu(i\hm_u) & i\hm_u J_{\nu+1}(i\hm_u)\\
 J_\nu(i\hm_d) & i\hm_d J_{\nu+1}(i\hm_d)
\end{array}\right|^2
}~.
\label{rho2connpq}
\eeq
We note that this result is of the structure expected from the replica limit
of the Toda lattice equation \cite{SplitV}.

\begin{figure*}[ht]
   \unitlength1.0cm
     \epsfig{file=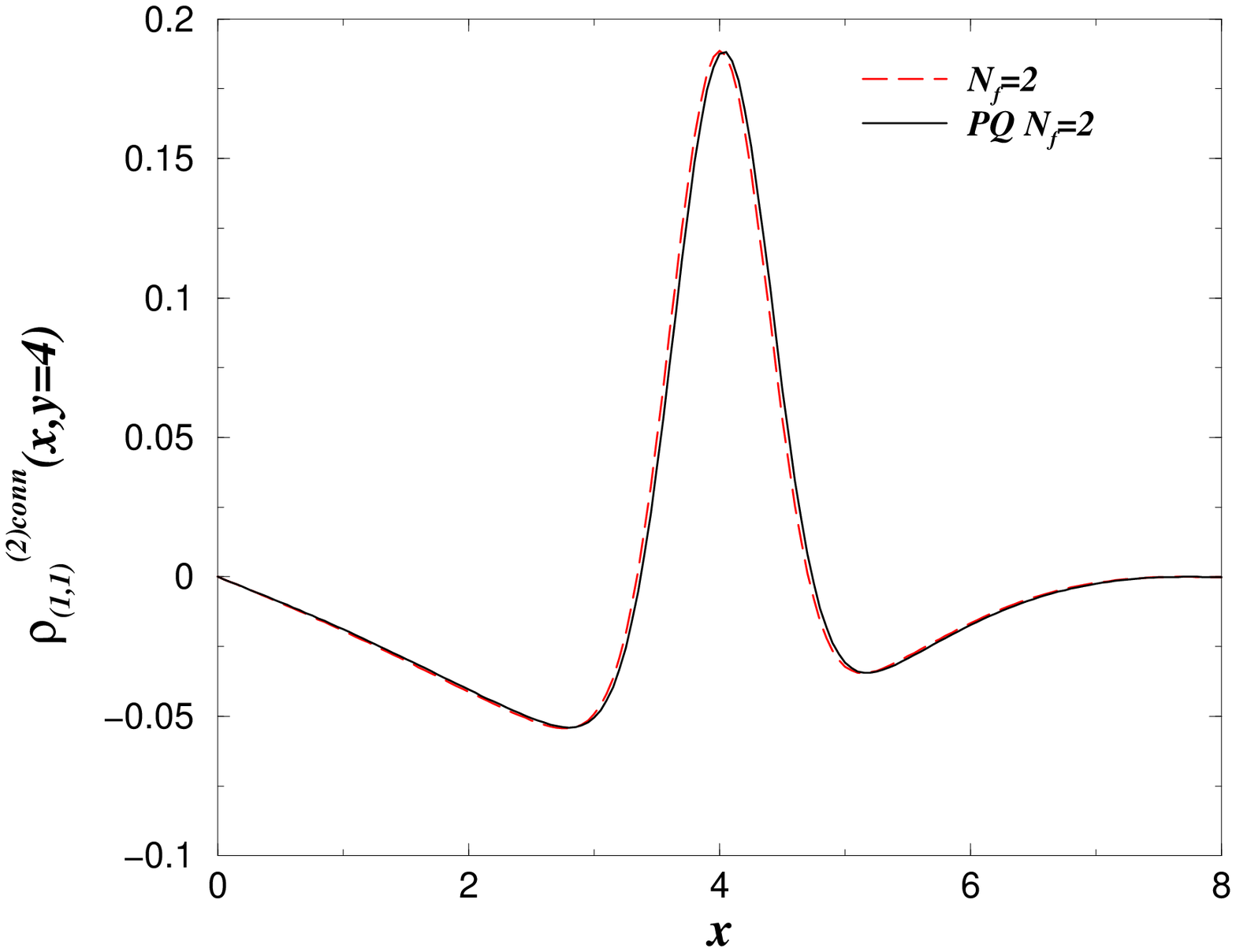,clip=,width=8cm}
     \epsfig{file=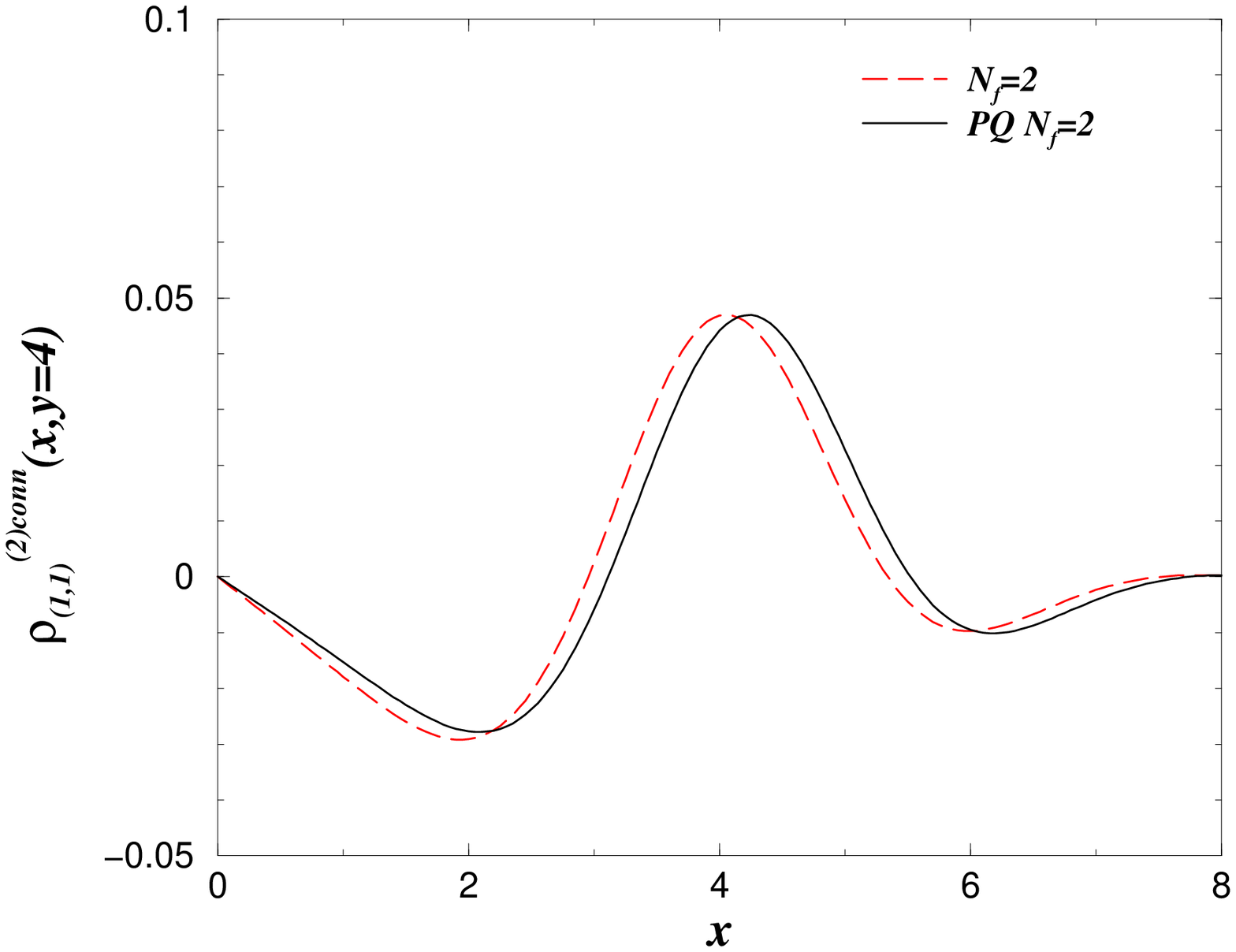,clip=,width=8cm}
     \caption{
   \label{fig:2} The partially quenched two point function (\ref{rho2connpq})
   compared to the unquenched two point function (\ref{rho2conn}). Here
   plotted as a
   function of the $D_1$ eigenvalue keeping the $D_2$ eigenvalue fixed
   at $\hat{y}=4$ and setting $\hm_u=3$ and $\hm_d=5$.
   The left hand figure is for $\hd=0.4$ and the
   right hand figure is for $\hd=1.0$. Notice the change of scale on
   the vertical axis.}
\end{figure*}

It is ideally suited for extracting $F_\pi$ from lattice simulations. To
visualize the effect of partially quenching we compare the partially quenched
two point function to that for $N_f=2$ in figure \ref{fig:2}.  
When the two masses are identical $m_u=m_d=m$ the expression for the partially
quenched two point function becomes
\beq
\rho_{(1,1)}^{(2)\,conn}(\hx,\hy) =
\hx\hy \ \frac{
\left|\begin{array}{ccc}
 \mathcal{I}^+(\hy,i\hm) & J_\nu(i\hm) & i\hm J_{\nu+1}(i\hm) \\
 \mathcal{I}^+(\hy,i\hm)' & J_\nu(i\hm)' & (i\hm J_{\nu+1}(i\hm))' \\
 \mathcal{I}^+(\hy,\hx)    & J_\nu(\hx)    &  \hx   J_{\nu+1}(\hx)
\end{array}\right|
\left|\begin{array}{ccc}
 -\mathcal{I}^0(\hx,i\hm) & J_\nu(i\hm) & i\hm J_{\nu+1}(i\hm) \\
 -\mathcal{I}^0(\hx,i\hm)' & J_\nu(i\hm)' & (i\hm J_{\nu+1}(i\hm))' \\
 \tilde{\mathcal{I}}^-(\hx,\hy) & e^{-\hd^2/2}J_\nu(\hy) &
 e^{-\hd^2/2}G_\nu(\hy,\hd)
\end{array}\right|
}{
\left|\begin{array}{cc}
 J_\nu(i\hm) & i\hm J_{\nu+1}(i\hm)\\
 J_{\nu-1}(i\hm) & i\hm J_{\nu}(i\hm)
\end{array}\right|^2
}~,
\label{rho2connpqdeg}
\eeq
where the prime indicates the derivative with respect to $i\hm$.

The two remaining building blocks are given as follows. From the 
definition (\ref{H}) we have from eq. (\ref{K2})
\beqn
H_{N+1}^{(2)}(x_1,x_2)
&=&\frac{h_{N+1}(x_1^2+m_u^2)(x_1^2+m_d^2)}{h_N^{(N_f)}(x_2^2+m_u^2)(x_2^2+m_d^2)}
\frac{\det\left[
\begin{array}{ccc}
H_{N+1}(x_1,im_u) & P_{N+1}(-m_u^2) & P_{N+2}(-m_u^2)\\
H_{N+1}(x_1,im_d) & P_{N+1}(-m_d^2) & P_{N+2}(-m_d^2)\\
H_{N+1}(x_1,x_2)    & P_{N+1}(x_2^2)    & P_{N+2}(x_2^2)\\
\end{array}
\right]}{\det\left[
\begin{array}{cc}
P_{N}(-m_u^2) & P_{N+1}(-m_u^2)\\
P_{N}(-m_d^2) & P_{N+1}(-m_d^2)\\
\end{array}
\right]} .\cr
&&
\label{H2}
\eeqn
The microscopic limit is readily obtained:
\beq
\lim_{N\to\infty}H_{N+1}^{(2)}(x_1,x_2) =
\frac{2N \hx_1^{\nu+1} (\hx_1^2+\hm_u^2) (\hx_1^2+\hm_d^2) }
 {\hx_2^{\nu}(\hx_2^2+\hm_u^2)(\hx_2^2+\hm_d^2)}
\frac{\det\left[
\begin{array}{ccc}
 \mathcal{I}^0(\hx_1,i\hm_u) & J_\nu(i\hm_u) & i\hm_u J_{\nu+1}(i\hm_u) \\
 \mathcal{I}^0(\hx_1,i\hm_d) & J_\nu(i\hm_d) & i\hm_d J_{\nu+1}(i\hm_d) \\
 \mathcal{I}^0(\hx_1,\hx_2)  & J_\nu(\hx_2) &   \hx_2 J_{\nu+1}(\hx_2)
\end{array}
\right]}{\det\left[
\begin{array}{cc}
J_\nu(i\hm_u) & i\hm_u J_{\nu+1}(i\hm_u)\\
J_\nu(i\hm_d) & i\hm_d J_{\nu+1}(i\hm_d)\\
\end{array}
\right]} .
\label{H2s}
\eeq
We note that it is $\hmu_f$-independent. 
From eq. (\ref{K2}) and the definition (\ref{Hh}) we have 
\beqn
\hH_{N+1}^{(2)}(y_1,y_2)
&=&\frac{h_{N+1}}{h_N^{(N_f)}}
\frac{\det\left[
\begin{array}{ccc}
K_{N+1}(y_1,im_u) & P_{N+1}(-m_u^2) & P_{N+2}(-m_u^2)\\
K_{N+1}(y_1,im_d) & P_{N+1}(-m_d^2) & P_{N+2}(-m_d^2)\\
\hH_{N+1}(y_1,y_2)    & \chi_{N+1}(y_2^2)    & \chi_{N+2}(y_2^2)\\
\end{array}
\right]}{\det\left[
\begin{array}{cc}
P_{N}(-m_u^2) & P_{N+1}(-m_u^2)\\
P_{N}(-m_d^2) & P_{N+1}(-m_d^2)\\
\end{array}
\right]} ,
\label{Hh2}
\eeqn
leading to 
\beq
\lim_{N\to\infty}\hH_{N+1}^{(2)}(y_1,y_2) =
2N \frac{\hy_2^{\nu+1}}{\hy_1^{\nu}}
\frac{\det\left[
\begin{array}{ccc}
 \mathcal{I}^+(\hy_1,i\hm_u) & J_\nu(i\hm_u) & i\hm_u J_{\nu+1}(i\hm_u) \\
 \mathcal{I}^+(\hy_1,i\hm_d) & J_\nu(i\hm_d) & i\hm_d J_{\nu+1}(i\hm_d) \\
 \mathcal{I}^0(\hy_1,\hy_2) & e^{-\hd^2/2} J_\nu(\hy_2) &
 e^{-\hd^2/2}G_\nu(\hy_2,\hd))
\end{array}
\right]}{\det\left[
\begin{array}{cc}
 J_\nu(i\hm_u) & i\hm_u J_{\nu+1}(i\hm_u)\\
 J_\nu(i\hm_d) & i\hm_d J_{\nu+1}(i\hm_d)
\end{array}
\right]} .
\label{Hh2s}
\eeq
The two densities are given by 
\beqn
\rho_{(1,0)}^{(2)}(\hx) &=& \lim_{N\to\infty} \frac{1}{2N} H_{N+1}^{(2)}(x,x)
\label{rho2x} \\
\rho_{(0,1)}^{(2)}(\hy) &=& \lim_{N\to\infty} \frac{1}{2N} \hH_{N+1}^{(2)}(y,y)\ .
\label{rho2y}
\eeqn
They differ because of the asymmetric flavor content, the $\hx$ eigenvalues
not coupling to $\hmu_2$. Eq. (\ref{rho2x}) agrees with the known result
\cite{DN}, as well as  eq. (\ref{rho2y}) when setting $\hmu_2=0$.

We are now ready to state the most general correlation function for $N_1=2$
and $N_2=0$ flavors. Upon using the above results, taking out common factors
of the determinant we obtain in terms of the abbreviations eq. (\ref{shortint})
\beqn
&&\rho_{(n,k)}^{(2)}(\{\hat{x}\}_n,\{\hat{y}\}_k) ~=~ \prod_{i=1}^n \hx_i\prod_{j=1}^k
\hy_j \det\left[
\begin{array}{cc}
 J_\nu(i\hm_u) & i\hm_u J_{\nu+1}(i\hm_u)\\
 J_\nu(i\hm_d) & i\hm_d J_{\nu+1}(i\hm_d)
\end{array}
\right]^{-n-k}\cr
&\times&
\left|
\begin{array}{cc}
\left|
\begin{array}{ccc}
 {\cal I}^0(\hx_{i_1},i\hm_u) & J_\nu(i\hm_u) & i\hm_u J_{\nu+1}(i\hm_u) \\
 {\cal I}^0(\hx_{i_1},i\hm_d) & J_\nu(i\hm_d) & i\hm_d J_{\nu+1}(i\hm_d) \\
 {\cal I}^0(\hx_{i_1},\hx_{i_2}) & J_\nu(\hx_{i_2}) &
 \hx_{i_2} J_{\nu+1}(\hx_{i_2})
\end{array}
\right|
&
\left|
\begin{array}{ccc}
 {\cal I}^0(\hx_{i_1},i\hm_u) & J_\nu(i\hm_u) & i\hm_u J_{\nu+1}(i\hm_u) \\
 {\cal I}^0(\hx_{i_1},i\hm_d) & J_\nu(i\hm_d) & i\hm_d J_{\nu+1}(i\hm_d) \\
 -\tilde{\cal I}^-(\hx_{i_1},\hy_{j_2}) & e^{-\hd^2/2} J_\nu(\hy_{j_2}) &
 e^{-\hd^2/2}G_\nu(\hy_{j_2},\hd)
\end{array}
\right|
\\
\\
\left|
\begin{array}{ccc}
 {\cal I}^+(\hy_{j_1},i\hm_u) & J_\nu(i\hm_u) & i\hm_u J_{\nu+1}(i\hm_u) \\
 {\cal I}^+(\hy_{j_1},i\hm_d) & J_\nu(i\hm_d) & i\hm_d J_{\nu+1}(i\hm_d) \\
 {\cal I}^+(\hy_{j_1},\hx_{i_2}) & J_\nu(\hx_{i_2}) &
 \hx_{i_2} J_{\nu+1}(\hx_{i_2})
\end{array}
\right|
&
\left|
\begin{array}{ccc}
 {\cal I}^+(\hy_{j_1},i\hm_u) & J_\nu(i\hm_u) & i\hm_u J_{\nu+1}(i\hm_u) \\
 {\cal I}^+(\hy_{j_1},i\hm_d) & J_\nu(i\hm_d) & i\hm_d J_{\nu+1}(i\hm_d) \\
 {\cal I}^0(\hy_{j_1},\hy_{j_2}) & e^{-\hd^2/2} J_\nu(\hy_{j_2}) &
 e^{-\hd^2/2}G_\nu(\hy_{j_2},\hd)
\end{array}
\right|
\end{array}
\right|.
\cr
&&
\label{allRnk2S}
\eeqn
As expected we verify that all correlation functions of eigenvalues $\hx_j$
are $\hmu$-independent and agree with the results of \cite{DN}.


\sect{The non-chiral Two-Matrix Problem}\label{sec-nonchiral}

Having discussed the chiral Unitary Two-Matrix Theory in great detail it is a small
step to go through the same analysis in the ordinary Unitary Two-Matrix Theory. As
mentioned briefly in the introduction, this ensemble is relevant for QCD in three
space-time dimensions, and may have applications in condensed matter physics as well \cite{HN}.

It is straightforward to consider two arbitrary chemical potentials $\mu_1$ and
$\mu_2$, as in the chiral case treated above. However, for simplicity we
restrict ourselves to just imaginary isospin chemical potential, $i.e.$ the case 
where $\mu_1 = -\mu_2 \equiv \mu$.
We present the most general case with an even number of flavors,
$2N_f=2N_++2N_-$. The techniques used here can also be applied to the
case of an odd number of flavors.  
The masses are pairwise equal in magnitude but of opposite sign 
\cite{VZ3,Sz}
\beq
 m_{+1},\ldots,m_{+N_+},-m_{+1},\ldots,-m_{+N_+},\ \ \mbox{and}\ \ 
m_{-1},\ldots,m_{-N_-},-m_{-1},\ldots,-m_{-N_-}.
\label{masscount}
\eeq
Both positive and negative mass fermions are coupled to imaginary
isospin potential. In intermediate steps also an odd number of flavors will
appear. 
The Dirac operator
then remains anti-hermitian as in four dimensions, and the corresponding Random
Two-Matrix Theory reads
\beqn
{\cal Z}^{(2N_f)}(\{m_+\},\{m_-\}) \sim
 \int\! d\Phi  d\Psi \exp\left[-{N}{\rm Tr}\left(\Phi^2
+\Psi^2\right)\right]
\prod_{f=1}^{2N_+} \det[{\cal D_+}(\mu) + m_{+f}]\!
\prod_{g=1}^{2N_-} \det[{\cal D_-}(\mu) + m_{-g}],
\label{ZNf3d}
\eeqn 
where both $\Phi$ and $\Psi$ now are $N\times N$ hermitian matrices, and
\beq
{\cal D}_{\pm}(\mu) ~\equiv~ i\Phi \pm i\mu\Psi ~\equiv~ i\Phi_\pm~.
\label{PhipmQCD3}
\eeq
A related non-hermitian matrix model aimed at real baryon chemical potential
was considered in ref. \cite{G3}. 
Here we again restrict ourselves to the Gaussian case
for convenience. The same remarks we made about non-Gaussian potentials from
the previous section also
apply here. For the Riemann-Hilbert problem to be solved in order to prove
universality in the quenched case we refer to \cite{BEH}.

We next perform the same change of variables as in eq. (\ref{Phipm}) to obtain
\beqn
{\cal Z}^{(2N_f)}(\{m_+\},\{m_-\})  \sim
 \int \!\! d\Phi_+  d\Phi_- \! \prod_{f=1}^{2N_+} \det[i\Phi_+ + m_{+f}]
\prod_{g=1}^{2N_-} \det[i\Phi_- + m_{-g}]
\exp\!\left[-\frac{N}{4}{\rm Tr}V(\Phi_+,\Phi_-)\right],
\label{ZNfpm3d}
\eeqn
with a potential
\beq
V(\Phi_+,\Phi_-) ~=~ \frac{\mu^2+1}{\mu^2}\left(\Phi_+^2 + \Phi_-^2\right)
+ 2\frac{\mu^2-1}{\mu^2}\Phi_+\Phi_- ~. \label{V3d}
\eeq
Going to an eigenvalue representation can be done exactly as in refs. \cite{IZ,M}.
Because of the pairing in positive and negative masses the integration measure
takes the following form
\beqn
{\cal Z}^{(2N_f)}(\{m_+\},\{m_-\})  &=& \int_{-\infty}^{\infty}
\prod_{i=1}^N\left(dx_idy_i \prod_{f=1}^{N_+}(x_i^2+m_{+f}^2) 
\prod_{g=1}^{N_-}(y_i^2+m_{-g}^2)\right) \cr
&&\hspace{-1cm}\times\Delta_N(\{x\})\Delta_N(\{y\})\det\left[\exp\left(N\frac{1-\mu^2}{2\mu^2}x_iy_j\right)\right] 
e^{-N\frac{\mu^2+1}{4\mu^2}\sum_i^N(x_i^2+y_i^2)} ~. \label{evrep3d}
\eeqn
Due to the presence of determinant
factors in the measure this is a generalization of the ordinary (quenched) Gaussian 
Unitary Random Two-Matrix Model
considered in the literature \cite{IZ,M,EM,E}. We seek biorthogonal polynomials satisfying
\beq
\int_{-\infty}^{\infty} dx dy \ w^{(2N_f)}(x,y)\ 
P_n^{(2N_f)}(x)\ 
Q_k^{(2N_f)}(y)
= h_n^{(2N_f)} \delta_{nk} ~, \label{bio3d}
\eeq
with respect to the weight function
\beq
w^{(2N_f)}(x,y)~\equiv~ 
\prod_{f=1}^{N_+}(x^2+m_{+f}^2)
\prod_{g=1}^{N_-}(y^2+{m}_{-g}^2)\ 
e^{N\frac{1-\mu^2}{2\mu^2}xy}e^{-N\frac{\mu^2+1}{4\mu^2}(x^2+y^2)}\ .
\label{weight3d}
\eeq
The only difference to the previous section is that the $I$-Bessel function is
replaced by an exponential and that we integrate over the full real line. 
Our polynomials are now polynomials in $x$ and $y$, and not in their squares. 
Furthermore all factors $(x^2+m_f^2)$ are counted as {\it two flavors} with
masses $\pm im_f$, instead of one in the previous section.

Taking into account these differences
all definitions of the transforms eq. (\ref{chidef}) and the
four kernels eqs. (\ref{K}) - (\ref{M}) carry trough and we shall not repeat
them here. The eigenvalue correlation functions defined as 
\beqn
R_{(n,k)}^{(2N_f)}(\{x\}_n,\{y\}_k) &\equiv&
\frac{N!^2}{(N-n)!(N-k)!}\frac{1}{{\cal Z}^{(2N_f)}(\{m_+\},\{m_-\})} 
\int_{-\infty}^{\infty} \prod_{j=n+1}^N dx_j \prod_{l=k+1}^N dy_l
\det\left[e^{N\frac{1-\mu^2}{2\mu^2}x_iy_j}\right]\cr
&\times&\prod_{i=1}^N\left( 
\prod_{f=1}^{N_+}(x_i^2+m_{+f}^2)\prod_{g=1}^{N_-}(y_i^2+m_{-g}^2)\right) 
 e^{-N\frac{\mu^2+1}{4\mu^2}\sum_i^N(x_i^2+y_i^2)}
 \Delta_N(\{x\})\Delta_N(\{y\})\ .
\cr
&&
\label{Revrep3d}
\eeqn
are given by the same formula eq. (\ref{allRnk}) in terms of the corresponding
non-chiral kernels.

\subsection{\sc The quenched case}

The quenched case corresponds to omitting the explicit factors of determinants
from the measure, while still performing the change of variables
(\ref{PhipmQCD3}). We then get the standard Random Two-Matrix Model, with a
coupling between the two matrices $\Phi_{\pm}$ as given by eq. (\ref{V3d}). 
We will need these results later when expressing
the unquenched results in terms of the quenched ones.

The biorthogonality condition (\ref{bio3d}) becomes
\beq
\int_{-\infty}^{\infty} dx dy \ 
e^{N\frac{1-\mu^2}{2\mu^2}xy}e^{-N\frac{\mu^2+1}{4\mu^2}(x^2+y^2)}P_n(x)P_m(y)
~=~ h_n \delta_{nm} ~. \label{bio3dQ}
\eeq
Because of the weight being symmetric in $x$ and $y$ we have that $Q_n(y)=P_n(y)$.
It was noted indirectly in refs. \cite{EK99} and 
\cite{MS} that this condition is related to Hermite polynomials,
although precise details were not given there. 
This can also be seen along the same lines as in appendix \ref{Lbiop}.
In our conventions we find for
the monic polynomials
\beq
P_n(x) ~=~ 2^{-n} \left(\frac{\mu^2+1}{N}\right)^{\frac{n}{2}}
H_n\left(\sqrt{\frac{N}{\mu^2+1}}\ x\right).
\label{op3d}
\eeq
For the normalization constants we obtain 
\beq
h_n ~=~ \frac{\mu\pi}{2^{n-1}N^{n+1}}\ n!\ (1-\mu^2)^{n}\ .
\label{norm3d}
\eeq
The factorization of the weight observed in Appendix \ref{wexp} also applies here. 
The weight $w(x,y)$ is given by the sum over the biorthogonal polynomials
\beqn
w(x,y) &=&  \exp\left[
  -N\frac{\mu^2+1}{4\mu^2}(x^2+y^2)+N\frac{1-\mu^2}{2\mu^2}xy\right]
\label{whh}
\\
&=& \frac{2\mu}{1+\mu^2}e^{-\frac{N}{\mu^2+1}(x^2+y^2)}
\sum_{k=0}^{\infty}\frac{1}{k!}  \left(\frac{1-\mu^2}{2(\mu^2+1)}\right)^k
H_k\left(\sqrt{\frac{N}{\mu^2+1}}\ x\right)H_k\left(\sqrt{\frac{N}{\mu^2+1}}\
y\right),\nn
\eeqn
using the Mehler formula for Hermite polynomials. The individual factors are
obviously $w_1(x)=w_2(x)=e^{-\frac{N}{\mu^2+1}x^2}$.

The functions $\chi_n(x)$ can be found explicitly by integrating the polynomials.
Alternatively, we could also use the results from the previous section by setting
$\nu=\pm\frac12$ and using the relation (\ref{HL}) below. 
The $I$-Bessel function then becomes a hyperbolic sine or
cosine and the two terms add up to an integral over the full real line.
The result is 
\beq
\chi_n(x) ~=~  \frac{\mu\sqrt{\pi}}{2^{n-1}N^{\frac{n+1}{2}}}
\frac{(1-\mu^2)^n}{(\mu^2+1)^{\frac{n+1}{2}}}
\ e^{-\frac{N}{\mu^2+1}x^2}
H_n\left(\sqrt{\frac{N}{\mu^2+1}}\ x\right)\ ,
\label{chi3d}
\eeq
again being proportional to the polynomials. 
The three different quenched kernels thus read as follows:
\beqn
K_N(y,x)&=& \sum_{k=0}^{N-1}\frac{N}{\pi\mu\ k!\
  2^{k+1}}\frac{(\mu^2+1)^k}{(1-\mu^2)^k}
H_k\left(\sqrt{\frac{N}{\mu^2+1}}\ x\right)H_k\left(\sqrt{\frac{N}{\mu^2+1}}\ y\right)
\label{K3d}\\
H_N(x,y)&=& \sum_{k=0}^{N-1}\frac{N^{\frac12}}{\sqrt{\pi}\ k!\
  2^{k}}\frac{1}{(\mu^2+1)^{\frac12}}
H_k\left(\sqrt{\frac{N}{\mu^2+1}}\ x\right)H_k\left(\sqrt{\frac{N}{\mu^2+1}}\ y\right)
\ e^{-\frac{N}{\mu^2+1}x^2}\cr
&=& \hH_N(y,x)
\label{H3d}\\
M_N(x,y)&=& \sum_{k=0}^{N-1}\frac{\mu}{k!\
  2^{k-1}}\frac{(1-\mu^2)^k}{(\mu^2+1)^{k+1}}
H_k\left(\sqrt{\frac{N}{\mu^2+1}}\ x\right)H_k\left(\sqrt{\frac{N}{\mu^2+1}}\ y\right)
e^{-\frac{N}{\mu^2+1}(x^2+y^2)}.
\label{M3d}
\eeqn
With this result we can now obtain all spectral correlation functions via eq. (\ref{allRnk}).

To take the scaling limit, we make use of the relations between Hermite and Laguerre
polynomials,
\beqn
H_{2n}(x) &=& (-1)^n2^{2n}n!\ L_n^{(-1/2)}(x^2) \cr
H_{2n+1}(x) &=& (-1)^n2^{2n+1}n!\ x\,L_n^{(1/2)}(x^2)\ ,
\label{HL}
\eeqn
and the asymptotic behavior of Laguerre polynomials (\ref{Lasymp}). This gives
\beqn
H_{2n}(X) &\to& (-1)^n 2^{2n}n!\ n^{-\frac12}\frac{1}{\sqrt{\pi}}\cos(\sqrt{4n}\ X) \cr
H_{2n+1}(X) &\to& (-1)^n 2^{2n+1}n!\ \frac{1}{\sqrt{\pi}}\sin(\sqrt{4n}\ X)\ .
\label{Hasympt}
\eeqn
If we rescale $\sqrt{2}Nx=\hx$,  $\sqrt{2}Ny=\hy$ leading to the customary
level spacing $\frac{1}{\pi}$
we arrive at the
following asymptotic for the kernels:
\beqn
\lim_{N\to\infty}K_N\left(x=\frac{\hx}{\sqrt{2}N},y=\frac{\hy}{\sqrt{2}N}\right) &=& 
\frac{N^2}{\sqrt{\pi}\hmu}\ 
{\cal I}^+(\hx,\hy) 
\label{Ks3d}\\
\lim_{N\to\infty}H_N\left(x=\frac{\hx}{\sqrt{2}N},y=\frac{\hy}{\sqrt{2}N}\right)&=& 
{\sqrt{2}\ N}\ {\cal I}^0(\hx,\hy)
\label{Hs3d} \\
&=& \lim_{N\to\infty}\hH_N\left(y=\frac{\hy}{\sqrt{2}N},x=\frac{\hx}{\sqrt{2}N}\right)
\cr
&& \cr
\lim_{N\to\infty}M_N\left(x=\frac{\hx}{\sqrt{2}N},y=\frac{\hy}{\sqrt{2}N}\right)
&=& 
2\hmu\sqrt{\pi}\ {\cal I}^-(\hx,\hy) \ ,
\label{Ms3d}
\eeqn
where we have introcued the following notation
\beqn
{\cal I}^\pm(x,y) &\equiv& \frac{1}{\pi}\int_0^1 ds\ 
e^{\pm\hat{\mu}^2s^2}\cos(s(x-y)) \cr 
{\cal I}^0(x,y) &\equiv& \frac{1}{\pi}\int_0^1 ds\ \cos(s(x-y))\ =\ \frac{1}{\pi}
\frac{\sin({x}-{y})}{({x}-{y})}\ .
\label{shortint3d}
\eeqn
The arguments can be real or imaginary. 
In the asymptotic we have replaced the sums by integrals and used again eq. (\ref{mut})
with $t=k/N$. For convenience we have kept the scaling $2N\mu^2=\hmu^2$
as in the previous section. 
The weight function has the following limit:
\beq
\lim_{N\to\infty}w\left(x=\frac{\hx}{\sqrt{2}N},y=\frac{\hy}{\sqrt{2}N}\right)~=~
\exp\left(-\frac{(\hx-\hy)^2}{8\hmu^2}\right). 
\label{wlim}
\eeq
We will also use the following abbreviation in this section:
\beq
\tilde{\cal I}^-(x,y) \ \equiv\ \frac{1}{2\hmu\sqrt{\pi}}
\exp\left(-\frac{(x-y)^2}{8\hmu^2}\right)-\frac{1}{\pi}\int_0^1 ds\ 
e^{-\hat{\mu}^2s^2}\cos(s(x-y)) \ . 
\eeq

Putting all together we obtain from the definition (\ref{microdef}) with
$(2N)^{-n-k}$ replaced by $(\sqrt{2}N)^{-n-k}$ 
\beq
\rho_{(n,k)}(\{\hx\},\{\hy\}) = \det\left[
\begin{array}{cc}
{\cal I}^0(\hx_{i_1},\hx_{i_2}) &
-\tilde{\cal I}^-(\hx_{i_1},\hy_{j_2})
\\
& \\
{\cal I}^+(\hy_{j_1},\hx_{i_2})
& {\cal I}^0(\hy_{j_1},\hy_{j_2})\\
\end{array}\right],
\label{allRnks3d}
\eeq
after taking out common factors from the determinant. As a check we obtain
that the following densities are constant and $\hmu$-independent:
\beq
\rho_{(1,0)}(\hx)~=~ \rho_{(0,1)}(\hx)~=~ \frac{1}{\pi} \ , 
\eeq
and we also recover the known correlations of the one-matrix model in terms of
the sine-kernel for $\rho_{(n,0)}({\hx})$  and  $\rho_{(0,k)}({\hy})$ respectively.


\subsection{\sc Two and more flavors}

In this subsection we outline how the insertion of flavors can be
reexpressed in terms of quenched quantities as in the last section, and we
will also give some examples. 

The biorthogonal polynomials enjoy again a determinental form as in eq. (\ref{charP})
\beqn
P_k^{(2N_f)}(x) &=& (-i)^k \langle \det[ {\mathcal{D_+}}(\mu) -ix ]\rangle_{{\cal
    Z}^{(2N_f)}} \cr
&=& \frac{(-)^k}{{\cal Z}^{(2N_f)}}  
 \int_{-\infty}^{\infty} \prod_{j=1}^k\left(dx_jdy_j (x_j - x)
\prod_{f}^{N_+} (x_j^2+m_{+f}^2)
\prod_{g}^{N_-}(y_j^2+{m}_{-g}^2)\right) \cr
&&\times\Delta_k(\{x\})\Delta_k(\{y\})\det\left[\exp(N\frac{1-\mu^2}{2\mu^2}x_iy_j)\right]\ 
e^{-N\frac{\mu^2+1}{4\mu^2}\sum_i^k(x_i^2+y_i^2)} \ ,
\label{charP3d}
\eeqn
and similarly for 
\beq
Q_k^{(2N_f)}(y) ~=~ (-i)^k \langle
\det {\mathcal{D_-}}(\mu) -iy \rangle_{{\cal Z}^{(2N_f)}}\ .
\eeq
The kernel of both polynomials is given by
\beqn
K_{N+1}^{(2N_f)}(y,x) &=& \frac{(-)^{k}}{h_N^{(2N_f)}} \langle \det[ {\mathcal{D_+}}(\mu) -i x ]
\det[ {\mathcal{D_-}}(\mu) -i y ]\rangle_{{\cal Z}^{(2N_f)}} \cr
&=&\frac{1}{h_N^{(2N_f)}{\cal Z}^{(2N_f)}}  
 \int_{-\infty}^{\infty} \prod_i^N\left(dx_idy_i (x_i - x)(y_i - y)
\prod_{f}^{N_+}(x_i^2+m_{+f}^2)
\prod_{g}^{N_-}(y_i^2+{m}_{-g}^2)\right) \cr
&&\times\Delta_N(\{x\})\Delta_N(\{y\})\det\left[\exp(N\frac{1-\mu^2}{2\mu^2}x_iy_j)\right]\ 
e^{-N\frac{\mu^2+1}{4\mu^2}\sum_i^N(x_i^2+y_i^2)} \ ,
\label{charK3d}
\eeqn
with the norms given in eq. (\ref{hNf}). 
We recall that when applying appendix \ref{allZ} all different masses in
eq. (\ref{masscount}) are counted as individual flavors, positive and
negative, leading to one column or row in eq. (\ref{Zgenresult}) each.
The remaining kernels can be
obtained from the definitions eqs. (\ref{H}) -  (\ref{M}). 
The most general correlation function then follows from eq. (\ref{allRnk}).

All these quantities, the transformed polynomials and the remaining kernels can
be in principle be expressed in terms of partition functions. 
They are given by
eq. (\ref{Zgenresult}): eqs. (\ref{ZP}), (\ref{ZK}), and by inserting 
the kernel, (\ref{ZK}), in the definitions (\ref{H}) -  (\ref{M}), with 
the obvious modification of dropping the $\nu$ terms and
writing polynomials with arguments of power one. 
However, there is a subtlety here as the partition function in
eq. (\ref{charP3d}) contains an odd number of flavors. There are two such
partition functions available \cite{VZ3,Sz}
\beqn
{\cal Z}_{\rm even}^{(2N_f+1)} &=& 
\int_{U(2N_{f}+1)}\! dU 
\cosh[V\Sigma {\mbox{\rm Tr}({\cal M}}U \Gamma_5 U^\dagger)]\ ,
\label{ZQCD3odd0}\\
{\cal Z}_{\rm odd}^{(2N_f+1)} &=& 
\int_{U(2N_{f}+1)}\! dU 
\sinh[V\Sigma {\mbox{\rm Tr}({\cal M}}U\Gamma_5 U^\dagger)]\ ,
\label{ZQCD3odd1}
\eeqn
with $\Gamma_5=$diag({\bf 1}$_{N_f}$,{\bf 1}$_{N_f+1})$, being even and odd in
the masses, respectively. Consequently the even 
polynomials $P_{2k}$ are given by eq. (\ref{ZQCD3odd0}) whereas the odd ones $P_{2k+1}$
are given by eq.  (\ref{ZQCD3odd1}). Similarly special care has to be taken
when adding single unpaired flavors to each of the sectors $N_+$ and $N_-$ 
as in the kernel eq.  (\ref{charK3d}), and we refer to \cite{Sz} for the
relation between partition functions of chiral and non-chiral theories at
$\mu=0$. 
Although similar massive partition functions with chemical potential have been 
computed from complex matrix models \cite{G3} and effective field theory 
\cite{SplitV} we do not further pursue this route. Alternatively we can
explicitly compute the large-$N$ limit from our biorthogonal polynomials and
kernels, as we will show in the following example.

Let us consider the simplest case with $2N_+=2$ flavors of masses  
$\pm m$, and no other flavors: $2N_-=0$. The corresponding weight function reads
\beq
w^{(2)}(x,y)~=~ 
(x^2+m^2)\ 
e^{N\frac{1-\mu^2}{2\mu^2}xy}e^{-N\frac{\mu^2+1}{4\mu^2}(x^2+y^2)}\ .
\eeq
The kernel is almost identical to the partially quenched one eq. (\ref{K2}):
\beqn
K_{N+1}^{(2)}(y,x)=\frac{h_{N+1}}{h_N^{(N_f)}(x-im)(x+im)}
\frac{\det\left[
\begin{array}{ccc}
K_{N+1}(y,im) & P_{N+1}(im) & P_{N+2}(im)\\
K_{N+1}(y,-im) & P_{N+1}(-im) & P_{N+2}(-im)\\
K_{N+1}(y,x)    & P_{N+1}(x^2)    & P_{N+2}(x^2)\\
\end{array}
\right]}{\det\left[
\begin{array}{cc}
P_{N}(im) & P_{N+1}(im)\\
P_{N}(-im) & P_{N+1}(-im)\\
\end{array}
\right]} .
\label{K23d}
\eeqn
The large-$N$ limit is easily obtained using the results of the previous
subsection, rescaling the mass as the eigenvalues $\sqrt{2}Nm=\hm$:
\beqn 
\lim_{N\to\infty}K_{N+1}^{(2)}(y,x)
&=&\frac{2N^2}{(\hx^2+\hm^2)}\ 
\frac{N^2}{\hmu\sqrt{\pi}}
\frac{\det\left[
\begin{array}{ccc}
{\cal I}^+(\hy,i\hm)
&\cosh(\hm) & -i\sinh(\hm)\\
{\cal I}^+(\hy,-i\hm)
&\cosh(\hm) & i\sinh(\hm)\\
{\cal I}^0(\hy,\hx)
&\cos(\hx) & -\sin(\hx)\\
\end{array}
\right]}{\det\left[
\begin{array}{cc}
i\sinh(\hm) & \cosh(\hm) \\
-i\sinh(\hm)& \cosh(\hm) \\
\end{array}
\right]} .
\label{K2s3d}
\eeqn
Here we have chosen $N$ to be even, noticing the difference between the
asymptotic of $P_N$ and $P_{N+2}$ by a factors of $(-1)$. Odd $N$ leads to the same result, as can
be seen from interchanging the last two columns in numerator and denominator.
Similarly the kernel can be seen to be real as it should. 
The remaining
kernels follow with the same modifications from eqs. (\ref{M2}), (\ref{H2})
and (\ref{Hh2}). Therefore  we just give the results here: 
\beqn 
\lim_{N\to\infty}M_{N+1}^{(2)}(x,y)
=2\hmu\sqrt{\pi}
\frac{(\hx^2+\hm^2)}{2N^{2}}
\frac{\det\left[
\begin{array}{ccc}
{\cal I}^0(\hy,i\hm)
&\cosh(\hm) & -i\sinh(\hm)\\
{\cal I}^0(\hy,-i\hm)
&\cosh(\hm) & i\sinh(\hm)\\
{\cal I}^-(\hx,\hy)
& e^{-\hat{\mu}^2}\cos(\hy) 
&  - e^{-\hat{\mu}^2}(\sin(\hy)+2\hat{\mu}^2\cos(\hy))\\
\end{array}
\right]}{\det\left[
\begin{array}{cc}
i\sinh(\hm)  &  \cosh(\hm)\\
-i\sinh(\hm) &\cosh(\hm) \\
\end{array}
\right]} ,
\label{M2s3d}\nn\\
\eeqn
\beqn 
\lim_{N\to\infty}H_{N+1}^{(2)}(x_1,x_2)
&=&
\sqrt{2}\ N \frac{(\hx_1^2+\hm^2)}{(\hx_2^2+\hm^2)}
\frac{\det\left[
\begin{array}{ccc}
{\cal I}^0(\hx_1,i\hm)
&\cosh(\hm) & -i\sinh(\hm)\\
{\cal I}^0(\hx_1,-i\hm)
&\cosh(\hm) & i\sinh(\hm)\\
{\cal I}^0(\hx_1,\hx_2)
&\cos(\hx_2) & -\sin(\hx_2)\\
\end{array}
\right]}{\det\left[
\begin{array}{cc}
i\sinh(\hm)  &  \cosh(\hm)\\
-i\sinh(\hm) &\cosh(\hm) \\
\end{array}
\right]} ,\cr 
&&
\label{H2s3d}
\eeqn
which is $\mu$-independent, and 
\beqn 
\lim_{N\to\infty}\hH_{N+1}^{(2)}(y_1,y_2)
=
\sqrt{2}\ N
\frac{\det\left[
\begin{array}{ccc}
{\cal I}^+(\hy_1,i\hm)
&\cosh(\hm) & -i\sinh(\hm)\\
{\cal I}^+(\hy_1,-i\hm)
&\cosh(\hm) & i\sinh(\hm)\\
{\cal I}^0(\hy_1,\hy_2)
& e^{-\hat{\mu}^2}\cos(\hy_2) 
&  -e^{-\hat{\mu}^2}(\sin(\hy_2)+2\hat{\mu}^2\cos(\hy_2))\\
\end{array}
\right]}{\det\left[
\begin{array}{cc}
i\sinh(\hm)  &  \cosh(\hm)\\
-i\sinh(\hm) &\cosh(\hm) \\
\end{array}
\right]} . \nn\\
\label{Hh2s3d}
\eeqn
We can write the general correlation function as follows:
\beqn
&&\rho_{(n,k)}^{(2)}(\{\hat{x}\}_n,\{\hat{y}\}_k) ~=~ 
\det\left[
\begin{array}{cc}
i\sinh(\hm)  &  \cosh(\hm)\\
-i\sinh(\hm) &\cosh(\hm) \\
\end{array}
\right]^{-n-k}\times\\
&&\hspace{-7mm}\left|
\begin{array}{cc}
\left|
\begin{array}{ccc}
{\cal I}^0(\hx_{i_1},i\hm)
&\cosh(\hm) & -i\sinh(\hm)\\
{\cal I}^0(\hx_{i_1},-i\hm)
&\cosh(\hm) & i\sinh(\hm)\\
{\cal I}^0(\hx_{i_1},\hx_{i_2})
&\cos(\hx_{i_2}) & -\sin(\hx_{i_2})\\
\end{array}
\right|
&
\left|
\begin{array}{ccc}
{\cal I}^0(\hx_{i_1},i\hm)
&\cosh(\hm) & -i\sinh(\hm)\\
{\cal I}^0(\hx_{i_1},-i\hm)
&\cosh(\hm) & i\sinh(\hm)\\
-\tilde{\cal I}^-(\hx_{i_1},\hy_{j_2})
&e^{-\hat{\mu}^2}\cos(\hy_{j_2}) 
& -e^{-\hat{\mu}^2}(\sin(\hy_{j_2})+2\hat{\mu}^2\cos(\hy_{j_2}))\\
\end{array}
\right|
\\
& \\
\left|
\begin{array}{ccc}
{\cal I}^+(\hy_{j_1},i\hm)
&\cosh(\hm) & -i\sinh(\hm)\\
{\cal I}^+(\hy_{j_1},-i\hm)
&\cosh(\hm) & i\sinh(\hm)\\
{\cal I}^+(\hy_{j_1},\hx_{i_2})
&\cos(\hx_{i_2}) & -\sin(\hx_{i_2})\\
\end{array}
\right|
&
\left|
\begin{array}{ccc}
{\cal I}^+(\hy_{j_1},i\hm)
&\cosh(\hm) & -i\sinh(\hm)\\
{\cal I}^+(\hy_{j_1},-i\hm)
&\cosh(\hm) & i\sinh(\hm)\\
{\cal I}^0(\hy_{j_1},\hy_{j_2})
&e^{-\hat{\mu}^2}\cos(\hy_{j_2}) 
& -e^{-\hat{\mu}^2}(\sin(\hy_{j_2})+2\hat{\mu}^2\cos(\hy_{j_2}))\\
\end{array}
\right|
\end{array}
\right|
\nn
\label{allRnk2S3d}
\eeqn
From the density 
\beq
\rho_{(1,0)}^{(2)}(\hx)=\lim_{N\to\infty}\frac{1}{\sqrt{2}N}H_{N+1}^{(2)}(x,x)
\eeq
and from $\rho^{(2)}_{(0,1)}(\hat{y})$ in the limit
$\hat{\mu} \to 0$ we recover the known massive density from the
second of ref. \cite{VZ3}. Similarly all higher
correlators at  $\hmu=0$ can be recovered.

Our second example is for $N_+=N_-=1$ flavors with masses $m_+$ and $m_-$. The structure is very similar
to the example in subsection \ref{2plus}, but with twice the flavor content:
$2N_f=2+2$ here. The four kernels at
finite-$N$ are similar to those given by eqs. (\ref{K11}), (\ref{M11}), (\ref{H11}) and
(\ref{Hh11}), with $2\times2$ and  $3\times3$ determinants in denominator and
numerator here. For example the first kernel reads
\newpage
\beqn 
K_{N+1}^{(2+2)}(y,x)&=&\frac{h_{N+1}}{h_N^{(N_f)}(x^2+m_+^2)(y^2+m_-^2)}\frac{1}{
\det\left[
\begin{array}{cc}
K_{N+1}(im_-,im_+) & K_{N+1}(im_-,-im_+)\\
K_{N+1}(-im_-,im_+) & K_{N+1}(-im_-,-im_+)\\
\end{array}
\right]
}\cr
&&\times 
\det\left[
\begin{array}{ccc}
K_{N+1}(im_-,im_+)  & K_{N+1}(im_-,-im_+)  &K_{N+1}(im_-,x)\\
K_{N+1}(-im_-,im_+) & K_{N+1}(-im_-,-im_+) &K_{N+1}(-im_-,x)\\
K_{N+1}(y,im_+)     & K_{N+1}(y,-im_+)     &K_{N+1}(y,x)\\
\end{array}
\right] , 
\label{K223d}
\eeqn
with an asymptotic limit 
\beqn
\lim_{N\to\infty}K_{N+1}^{(2+2)}(y,x)&=&\frac{2N^2}{(\hx^2+\hm_+^2)(\hy^2+\hm_-^2)}
\frac{N^2}{\sqrt{\pi}\hmu}
\frac{1}{
\det\left[
\begin{array}{cc}
{\cal I}^+(i\hm_-,i\hm_+) & {\cal I}^+(i\hm_-,-i\hm_+)\\
{\cal I}^+(-i\hm_-,i\hm_+) & {\cal I}^+(-i\hm_-,-i\hm_+)\\
\end{array}
\right]
}\cr
&&\times 
\det\left[
\begin{array}{ccc}
{\cal I}^+(i\hm_-,i\hm_+)  & {\cal I}^+(i\hm_-,-i\hm_+)  &{\cal I}^+(i\hm_-,\hx)\\
{\cal I}^+(-i\hm_-,i\hm_+) & {\cal I}^+(-i\hm_-,-i\hm_+) &{\cal I}^+(-i\hm_-,\hx)\\
{\cal I}^+(\hy,i\hm_+)     & {\cal I}^+(\hy,-i\hm_+)     &{\cal I}^+(\hy,\hx)\\
\end{array}
\right] . 
\label{K223ds}
\eeqn
The remaining kernels can be obtained similary, and we only give their
asymptotic expressions:
\beqn
\lim_{N\to\infty}M_{N+1}^{(2+2)}(x,y)&=&\frac{(\hx^2+\hm_+^2)(\hy^2+\hm_-^2)}{2N^2}
2\sqrt{\pi}\hmu\frac{1}{
\det\left[
\begin{array}{cc}
{\cal I}^+(i\hm_-,i\hm_+) & {\cal I}^+(i\hm_-,-i\hm_+)\\
{\cal I}^+(-i\hm_-,i\hm_+) & {\cal I}^+(-i\hm_-,-i\hm_+)\\
\end{array}
\right]
}\cr
&&\times 
\det\left[
\begin{array}{ccc}
{\cal I}^+(i\hm_-,i\hm_+)  & {\cal I}^+(i\hm_-,-i\hm_+)  &{\cal I}^0(i\hm_-,\hy)\\
{\cal I}^+(-i\hm_-,i\hm_+) & {\cal I}^+(-i\hm_-,-i\hm_+) &{\cal I}^0(-i\hm_-,\hy)\\
{\cal I}^0(\hx,i\hm_+)     & {\cal I}^0(\hx,-i\hm_+)     &{\cal I}^-(\hx,\hy)\\
\end{array}
\right] , 
\label{M223ds}\\
&& \cr
\lim_{N\to\infty}H_{N+1}^{(2+2)}(x_1,x_2)&=&\frac{(\hx_1^2+\hm_+^2)}{(\hx_2^2+\hm_-^2)}
\sqrt{2}\ N
\frac{1}{
\det\left[
\begin{array}{cc}
{\cal I}^+(i\hm_-,i\hm_+) & {\cal I}^+(i\hm_-,-i\hm_+)\\
{\cal I}^+(-i\hm_-,i\hm_+) & {\cal I}^+(-i\hm_-,-i\hm_+)\\
\end{array}
\right]
}\cr
&&\times 
\det\left[
\begin{array}{ccc}
{\cal I}^+(i\hm_-,i\hm_+)  & {\cal I}^+(i\hm_-,-i\hm_+)  &{\cal I}^+(i\hm_-,\hx_2)\\
{\cal I}^+(-i\hm_-,i\hm_+) & {\cal I}^+(-i\hm_-,-i\hm_+) &{\cal I}^+(-i\hm_-,\hx_2)\\
{\cal I}^0(\hx_1,i\hm_+)     & {\cal I}^0(\hx_1,-i\hm_+)     &{\cal I}^0(\hx_1,\hx_2)\\
\end{array}
\right] , 
\label{H223ds}\\
&& \cr
\lim_{N\to\infty}\hH_{N+1}^{(2+2)}(y_1,y_2)&=&\frac{(\hy_2^2+\hm_-^2)}{(\hy_1^2+\hm_+^2)}
\sqrt{2}\ N
\frac{1}{
\det\left[
\begin{array}{cc}
{\cal I}^+(i\hm_-,i\hm_+) & {\cal I}^+(i\hm_-,-i\hm_+)\\
{\cal I}^+(-i\hm_-,i\hm_+) & {\cal I}^+(-i\hm_-,-i\hm_+)\\
\end{array}
\right]
}\cr
&&\times 
\det\left[
\begin{array}{ccc}
{\cal I}^+(i\hm_-,i\hm_+)  & {\cal I}^+(i\hm_-,-i\hm_+)  &{\cal I}^0(i\hm_-,\hy_2)\\
{\cal I}^+(-i\hm_-,i\hm_+) & {\cal I}^+(-i\hm_-,-i\hm_+) &{\cal I}^0(-i\hm_-,\hy_2)\\
{\cal I}^+(\hy_1,i\hm_+)     & {\cal I}^+(\hy_1,-i\hm_+)     &{\cal I}^0(\hy_1,\hy_2)\\
\end{array}
\right] . 
\label{Hh223ds}
\eeqn
The general correlation function
$\rho_{(n,k)}^{(2+2)}(\{\hat{x}\}_n,\{\hat{y}\}_k)$ can be easily obtained from the above. As a
check we have again compared the density 
\beq
\rho_{(1,0)}^{(2+2)}(\hx)=\lim_{N\to\infty}\frac{1}{\sqrt{2}N}H_{N+1}^{(2+2)}(x,x)
\eeq
in the limit $\hmu\to0$ to the known result \cite{VZ3}, finding agreement for
nondegenerate $\hm_+$ and  $\hm_-$.

\sect{Summary}\label{sec-summary}

We have introduced a chiral Two-Matrix Gaussian Unitary Ensemble in
Random Matrix Theory, and solved it explicitly both at finite matrix
size $N$ and in the $N \to \infty$ scaling limit. The scaling theory
describes the Dirac eigenvalue correlations of QCD in the
$\epsilon$-regime when subjected to imaginary isospin chemical
potential. In QCD language the first non-trivial mixed correlation
function has a strong dependence on the pion decay constant
$F_{\pi}$. This makes it possible to use such correlation functions to
extract $F_{\pi}$ from lattice gauge theory simulations
\cite{DHSS,DHSST}. Previously two special cases had been considered
from the viewpoint of effective field theory.  Here using the Random
Matrix Theory formulation we have computed the general correlation
functions. We have confirmed that in the two special cases ($N_f=0,2$)
our general expressions from the present Random Matrix Theory
calculation agree with the computations from effective field theory
\cite{DHSS,DHSST}.

A particularly useful by-product of the present work is the
``partially quenched'' case, where the Dirac operators that define the
theory do no couple to chemical potentials.  One can still measure the
eigenvalue correlations of the Dirac operators $D_1$ and $D_2$ on such
ensembles, and they will again show a very non-trivial dependence on
the pion decay constant $F_{\pi}$. From the point of view of a
practical implementation in lattice gauge theory simulations this is
the ideal set-up. The corresponding calculation in the effective field
theory seems technically very difficult without some new short-cut or
trick. Here the Random Matrix Theory provides a novel result that is
ready to be used to determine the pion decay constant $F_{\pi}$ from
lattice simulations.

We have also explicitly found the finite-$N$ solution for the ordinary
Two-Matrix Unitary Ensemble including mass terms. This extends the existing
literature. Our results for the corresponding scaling limit are also 
new. These results are of relevance for QCD in 3 space-time
dimensions. We have also shown directly how to recover the ordinary
One-Matrix results from the Two-Matrix theory.

\vspace{0.4cm}
\noindent
{\sc Acknowledgement:} We thank U. Heller, E. Kanzieper, B. Svetitsky,
D. Toublan and J. Verbaarschot for discussions, 
and we in particular thank B. Svetitsky for suggesting a computation of the
partially quenched case. We are grateful to D. Toublan for having explored an initial 
calculation for that problem in the effective field theory framework.

The work of G.A. and P.H.D. was partly supported by the 
European Community Network ENRAGE  (MRTN-CT-2004-005616).
The work of G.A was also supported by EPSRC grant EP/D031613/1. 
K.S. was supported by the Carlsberg Foundation. J.C.O. was supported in part
by U.S. DOE grant DE-FC02-01ER41180. 

\begin{appendix}

\sect{Biorthogonal Laguerre polynomials}\label{Lbiop}

We wish to show here that for $N_f=0$ the integral representation eq. (\ref{charP})
immediately leads to the fact that the quenched biorthogonal polynomials are
given by Laguerre polynomials as stated in eq. (\ref{polP}). To that aim we
will integrate out one of the matrices being Gaussian to map the polynomial to
the standard matrix representation of the orthogonal Laguerre
polynomials in the Gaussian chiral {\it one}-matrix theory. 
There the following fact is known:
\beqn
p_k(x^2) &=& (-)^k(ix)^{-\nu}\langle \det({\cal D} +ix) \rangle 
         ~\sim~ \int d\Phi \det\left[ 
\begin{array}{cc}
ix & i\Phi \\
i\Phi^\dag &ix \\
\end{array}
\right]
\exp[-N{\rm Tr}\Phi^\dag\Phi]
\label{p1MM} \\
&\sim& \int_0^\infty \prod_{j=1}^k \left( dx_j^2 x_j^{2\nu} (x^2- x_j^2) \right)
\exp\left[-N\sum_{i=1}^kx_i^2\right] \Delta_k(\{x^2\})^2  \ , \nonumber
\eeqn
leading to 
\beq
p_k(x^2) ~\sim~ L_k^{(\nu)}(Nx^2) \ 
\eeq
as orthogonal polynomials with respect to the weight $w(x)=
x^{2\nu+1}\exp[-Nx^2]$. Here we have ommited all normalisation factors. The
matrix $\Phi$ is complex and of size $k\times (k+\nu)$.

This can be compared to the representation of our biorthogonal polynomials
eq. (\ref{charP}): 
\beq
P_k(x^2) ~=~  (-)^k(ix)^{-\nu}\frac{1}{{\cal Z}_\nu} \int d\Phi_1d\Phi_2
 \det({\cal D}_1+ix) \exp\left[ -N V(\Phi_1,\Phi_2)\right]
\eeq
with the potential given in eq. (\ref{V}).
Since $P_k(x^2)$ is given by the
expectation value of a determinant depending on only {\it one} of the matrices, $\Phi_1$, the matrix
integration over $\Phi_2$ decouples. Completing the square and integrating out $\Phi_2$
we obtain an expectation value only with respect to  $\Phi_1$, with potential
\beq
V(\Phi_1) ~=~ (c_1~-~d^2/c_2) ~ {\rm Tr}\Phi_1^\dag\Phi_1
 ~=~ \frac{1}{1+\mu_1^2} ~ {\rm Tr}\Phi_1^\dag\Phi_1 \ .
\eeq 
Thus we have arrived exactly at eq. (\ref{p1MM}) with $N \to N/(1+\mu_1^2)$.
Consequently our biorthogonal polynomials are Laguerre polynomials as
was claimed in eq. (\ref{polP}). The same argument immediately leads to 
$Q_k(y^2)$ being Laguerre polynomials as well, integraing out $\Phi_1$. 

It is interesting to compare this to the orthogonality of the Laguerre
 polynomials in the complex plane which arises in the solution of the
 chiral two-matrix theory for a real chemical potential \cite{O}.
A proof of the orthogonality was given in \cite{G2} by direct computation.
An alternative proof can be formulated since the orthogonal polynomials in
 the complex plane also enjoy a determinental representation.
Even though the determinant contains {\it both} matrices of the two-matrix
theory one can still find a way to shift the variables of integration such
that one of the matrices can be easily integrated out leaving a form that
again leads to Laguerre polynomials \cite{ADOS2}.
However the expansion method (see Appendix \ref{wexp}) can not be
directly applied since the biorthogonal polynomials $p_k(z)$ and
 $p_k^*(z)$ are not functions of independent variables.



\sect{Biorthogonal polynomials from the expansion of the weight}\label{wexp}

Consider two sets of polynomials, $p_k(x)$ and $q_k(y)$, orthogonal with
 respect to the two weights $w_1(x)$ and $w_2(y)$ respectively
\beqn
\int dx ~ w_1(x) ~ p_k(x) ~ p_\ell(x) &=& f_k ~ \delta_{k,\ell} \nn \\
\int dy ~ w_2(y) ~ q_k(y) ~ q_\ell(y) &=& g_k ~ \delta_{k,\ell} ~.
\eeqn
Then it follows automatically that $p_k(x)$ and $q_k(y)$ are
 {\it bi}-orthogonal with respect to the weight $w(x,y)$ defined as follows
\beqn
w(x,y) &=& w_1(x) ~ w_2(y) ~
 \sum_{k=0}^\infty ~ \frac{h_k}{f_k g_k} ~ p_k(x) ~ q_k(y) ~.
\label{wexp-w}
\eeqn
The corresponding biorthogonality relation is
\beqn
\int dx \, dy ~ w(x,y) ~ p_k(x) ~ q_\ell(y) &=& h_k ~ \delta_{k,\ell} ~.
\label{wexp-biop}
\eeqn
While not all weights can be written in the form (\ref{wexp-w})
we will show below that the quenched weight
(given in (\ref{weight}) with $N_f=0$) in fact can.

Using these conventions one can easily obtain the transformed functions
\beqn
\chi_k(y) ~~=& \int dx ~ w(x,y) ~ p_k(x) &=~~ (h_k/g_k) ~ w_2(y) ~ q_k(y) \nn\\
\hchi_k(x) ~~=& \int dy ~ w(x,y) ~ q_k(y) &=~~ (h_k/f_k) ~ w_1(x) ~ p_k(x)
\label{wexp-bt}
\eeqn
and all the kernels
\beqn
K_N(y,x) &=& \sum_{k=0}^{N-1} \frac{1}{h_k} ~ q_k(y) ~ p_k(x) \nn\\
H_N(x_1,x_2) &=& w_1(x_1) \sum_{k=0}^{N-1} \frac{1}{f_k} ~ p_k(x_1) ~ p_k(x_2) \nn\\
\hH_N(y_1,y_2) &=& w_2(y_2) \sum_{k=0}^{N-1} \frac{1}{g_k} ~ q_k(y_1) ~ q_k(y_2) \nn\\
M_N(x,y) &=& w_1(x) ~ w_2(y) \sum_{k=0}^{N-1} \frac{h_k}{f_k g_k} ~ p_k(x) ~ q_k(y) ~.
\label{wexp-kern}
\eeqn
Interestingly we see that the weight function has the same form as the $M$
kernel just with different limits in the sum.  We can then write one of the
terms appearing in generalized ``Dyson Theorem'' (\ref{allRnk}) as
(for $N_f=0$)
\beq
w(x,y) - M_N(x,y) ~=~  w_1(x) ~ w_2(y) ~
 \sum_{k=N}^\infty ~ \frac{h_k}{f_k g_k} ~ p_k(x) ~ q_k(y) ~.
\label{w-M}
\eeq

In order to apply this to the current problem we make use of the known
 expansion of $I_\nu$ in terms of Laguerre polynomials giving (for $N_f=0$)
\beqn
w^{(0)}(x,y) &=& (xy)^{\nu+1} ~ e^{-N(c_1 x^2+ c_2 y^2)} ~
  I_{\nu}\left(2 d N x y\right) \label{B7}\\
 &=& (N d)^\nu \tau^{\nu+1} (x y)^{2\nu+1} e^{-N\tau (c_1 x^2 + c_2 y^2)}
 \sum_{n=0}^{\infty} \frac{n!(1-\tau)^n}{(n+\nu)!} 
 L_n^{(\nu)} (N \tau c_1 x^2) L_n^{(\nu)} (N \tau c_2 y^2) \nn
\eeqn
with $\tau = 1-d^2/(c_1 c_2)$.
Defining the monic polynomials
\beqn
p_k(x) &=& k! (-N \tau c_1)^{-k} L_k^{(\nu)} (N \tau c_1 x^2) \nn\\
q_k(y) &=& k! (-N \tau c_2)^{-k} L_k^{(\nu)} (N \tau c_2 y^2)
\eeqn
and the weights
\beqn
w_1(x) &=& x^{2\nu+1} e^{-N\tau c_1 x^2} \nn\\
w_2(y) &=& y^{2\nu+1} e^{-N\tau c_2 y^2}
\label{wexp-wop}
\eeqn
with support on the positive real line we get the normalisation of the polynomials
\beqn
f_k &=& \frac{k! (k+\nu)!}{2 (N\tau c_1)^{2k+\nu+1}} \nn\\
g_k &=& \frac{k! (k+\nu)!}{2 (N\tau c_2)^{2k+\nu+1}} \ .
\label{wexp-nop}
\eeqn
From eq. (\ref{B7}) we can then read off the norm of the biorthogonal polynomials 
\beqn
h_k &=& \frac{k! (k+\nu)! (N d)^{2k+\nu}}{4 (N^2\tau c_1 c_2)^{2k+\nu+1}} ~.
\label{wexp-hn}
\eeqn
The same considerations also apply to the non chiral case in terms of hermite
polynomials, cf. eq. (\ref{whh}).


\sect{All partition functions}\label{allZ}

In this appendix we derive the most general partition function, containing a
different number of $N_1$ flavors with masses $m_{f1}$ and $N_2$ flavors with
masses $m_{f2}$, where $N_f=N_1+N_2$. This result
is particularly useful as both the unquenched polynomials $P_k^{(N_f)}$ and  $Q_k^{(N_f)}$
as well as their kernel can be expressed in terms of these partition
functions, see eqs. (\ref{ZP}) and (\ref{ZK}).

Following the
same steps as in section \ref{ch2MM} the partition function 
has the following eigenvalue representation:
\beqn
{\cal Z}_{\nu}^{(N_f)}(\{m_1\},\{m_2\}) &=& {\cal N}(\{m_{1,2}\})
 \int_0^{\infty} \prod_{i=1}^N\left( dx_i dy_i (x_i y_i)^{\nu+1}
 \prod_{f=1}^{N_1}(x_i^2+m_{f1}^2)\prod_{f=1}^{N_2}(y_i^2+m_{f2}^2)\right) \cr
&\times& \Delta_N(\{x^2\})\Delta_N(\{y^2\})
 \det[I_{\nu}(2dN x_i y_j)]
 e^{-N\sum_i(c_1 x_i^2 + c_2 y_i^2)} ~.
\label{evrepgen}
\eeqn
Because all eigenvalues are integrated out the symmetry of the Vandermonde
determinants can be used to bring all $N!$ terms in the sum over permutations from 
the determinant of the $I$-Bessel function into diagonal form\footnote{This is
no longer true in correlation functions $R_{n,k}$ when less than $N$ of each
eigenvalues are integrated out.}
\beqn
{\cal Z}_{\nu}^{(N_f)}(\{m_1\},\{m_2\}) &=& {\cal N}(\{m_{1,2}\}) N!
 \int_0^{\infty} \prod_{i=1}^N\left( dx_i dy_i (x_i y_i)^{\nu+1}
 \prod_{f=1}^{N_1}(x_i^2+m_{f1}^2)\prod_{f=1}^{N_2}(y_i^2+m_{f2}^2)\right) \cr
&\times& \Delta_N(\{x^2\})\Delta_N(\{y^2\})
 \prod_{i=1}^{N}I_{\nu}(2dN x_i y_i)~
 e^{-N\sum_i(c_1 x_i^2 + c_2 y_i^2)} ~.
\label{evrepgenN}
\eeqn
Thus we can now interpret the  $I$-Bessel function as being part of the
weight. In this form we can apply a theorem that was proven in \cite{AV} for
matrix theories with complex eigenvalues, expressing such partition
functions (also called characteristic polynomials) in terms of the quenched
polynomials. We can apply this result by formally identifying
$z\rightarrow x$ and $z^* \rightarrow y$, thus relating our polynomials
to the polynomials ${\cal P}$ there as follows:
${\cal P}_k(z^2)\rightarrow P_k(x^2)$
and ${\cal P}_k(z^{2})^*\rightarrow Q_k(y^2)$.
Note that unlike in \cite{AV}, where the two sets of polynomials are related
by complex conjugation, our polynomials $P$ and $Q$ are no longer related by a
symmetry in general. However, this property was not used in the proof of the
theorem we wish to apply.

The quenched kernel made of the monic polynomials without the weight, 
\beq
K_{N}(y,x) ~=~ \sum_{k=0}^{N-1} \frac{Q_k(y^2)P_k(x^2)}{h_k} \ , 
\label{preK}
\eeq
is obviously self-contractive, due to the biorthogonality:
\beq
\int_0^\infty dxdy\ w^{(0)}(x,y)~K_{N}(y,u)K_{N}(v,x) ~=~ K_{N}(v,u)\ .
\eeq
This makes Dyson's
Theorem for integrating over determinants of such kernels applicable, which was
used in the derivation in \cite{AV}. We therefore arrive at 
\beqn
{\cal Z}_{\nu}^{(N_f)}(\{m_1\},\{m_2\}) &=& \prod_f^{N_1}m_{f1}^\nu \prod_f^{N_2}m_{f2}^\nu
\frac{ N! \prod_{j=0}^{N+N_2-1}h_j}{\Delta_{N_1}(\{m_1^2\})\Delta_{N_2}(\{m_2^2\})}\cr
&&\times
\det\left[ K_{N+N_2}(im_{l2},im_{k1}) \ P_{N+N_2}((im_{k1})^2)\ldots P_{N+N_1-1}((im_{k1})^2)\right],
\label{Zgenresult}
\eeqn
where we have given the result for $N_1\geq N_2$. The first $N_2$ columns are
given by  kernels and the last $N_1-N_2$ columns by monic polynomials of
increasing degree. The corresponding result for $N_2>N_1$ then obviously
follows by exchanging $N_1\leftrightarrow N_2$ and $m_{f1}\leftrightarrow m_{f2}$ 
in the above formula. 

The fact that imaginary arguments appear is related to the fact that mass terms
are positive, $\prod_i(x_i^2+m_1^2)$. Polynomials and kernels of real
arguments would appear, if we had considered characteristic polynomials, such
as  $\prod_i(x_i^2-u^2)$.

\end{appendix}

\end{document}